\documentclass[twocolumn, twocolappendix, trackchanges]{aastex701}
\usepackage{graphicx} 
\usepackage{xspace}
\usepackage{framed} 
\usepackage{amsmath}
\usepackage{txfonts}
\usepackage{rotating}
\usepackage{ulem}
\usepackage{booktabs}
\usepackage{subfigure}
\usepackage{float}
\usepackage{enumitem}
\usepackage{amssymb}
\usepackage[dvipsnames]{xcolor}
\usepackage{sistyle}
\SIthousandsep{,}
\usepackage{natbib}
\usepackage{multirow}
\usepackage{array}
\newcolumntype{C}{>{\centering\arraybackslash}m{0.42cm}}

\defcitealias{springel_03}{SH03}
\defcitealias{ploeckinger_20}{PS20}
\newcommand{\sh}{\citetalias{springel_03}\xspace}

\newcommand{\arepo}{\textsc{Arepo}\xspace}

\newcommand{\art}{\textsc{ART}\xspace}

\newcommand{\mmh}{M_{\rm max,\, hot}}
\newcommand{\tcool}{\eta_{\rm cool}}
\newcommand{\nh}{\langle n_{\rm H}\rangle}

\begin{document}

\title{Shock-heated Away: The Impact of Radiative Cooling on Gas-Phase Transitions in Supernova Remnants}

\author[orcid=0009-0003-0415-404X,sname='Zuzanna Kocjan']{Zuzanna Kocjan}
\affiliation{Department of Astronomy, University of Maryland, College Park, MD 20742, USA}
\email[show]{zkocjan@umd.edu}  

\author[orcid=0000-0001-9568-7287,sname='Benedikt Diemer']{Benedikt Diemer}
\affiliation{Department of Astronomy, University of Maryland, College Park, MD 20742, USA}
\email[show]{diemer@umd.edu}  

\author[orcid=0000-0002-6648-7136,sname='Vadim Semenov']{Vadim A. Semenov}
\affiliation{Center for Astrophysics, Harvard \& Smithsonian, 60 Garden St, Cambridge, MA 02138, USA}
\email[]{} 

\author[orcid=0000-0002-0404-003X,sname='Shmuel Bialy']{Shmuel Bialy}
\affiliation{Technion - Israel Institute of Technology, Haifa, 3200003 Israel}
\email[]{}  

\author[orcid=0000-0003-3948-6813, sname='Uri Malamud']{Uri Malamud}
\affiliation{Technion - Israel Institute of Technology, Haifa, 3200003 Israel}
\email[]{}  

\begin{abstract}

Supernova (SN) feedback plays a central role in regulating the structure of the interstellar medium (ISM) through the injection of energy and momentum. The amount of hot gas produced by a supernova is a key quantity that determines how efficiently SN feedback heats the ISM and drives mass exchange between its different gas phases, here defined as cold ($T < 10^3$ K), warm ($10^3\, \mathrm{K} < T < 2\times10^4\, \rm K$), and hot ($T > 2\times10^4$ K) gas. However, previous studies have reported discrepant amounts of hot gas formed under otherwise similar ambient conditions. To resolve these disagreements, we quantify the amount of hot gas produced by individual SN explosions using a suite of controlled simulations spanning a broad range of ISM environments that include both uniform and turbulent, multiphase backgrounds. We show that radiative cooling is a key factor regulating hot-gas production, and that differences in cooling efficiency can account for some of the discrepancies reported in the literature. We derive a simple predictive relation for the peak hot-gas mass attained during the evolution of a supernova remnant in terms of the mean ambient density, the initial phase distribution, and the efficiency of gas cooling, which we parameterize as the cooling time over a key temperature range of $10^{4.5}\mathrm{K} \lesssim T \lesssim 10^{5.1}\mathrm{K}$. Finally, using tracer particles, we distinguish the evaporation of cold and warm gas into the hot phase and derive physically motivated expressions for the evaporation efficiency. Our results provide simple, predictive relations for hot-gas production and phase transitions that can be incorporated into subgrid models of SN feedback in galaxy formation simulations.

\end{abstract}


\section{Introduction} \label{sec:intro}

Supernovae are widely recognized as major drivers of the thermodynamic and dynamical evolution of the interstellar medium (ISM), injecting energy and momentum into the surrounding gas and driving the redistribution of mass and energy across different gas phases, including the production of hot ionized gas (e.g., \citealt{1998_thornton, martizzi_15, kim_15, steinwandel_20, guo2024}). Their influence spans scales from the microphysics of cloud evaporation and phase exchange \citep{1977_cowie, 1977_mckee_cowie, mckee_77, 1991_white_long} and the production of X-ray-emitting shocked gas \citep{hamilton_x-ray_1983, borkowski_supernova_2001, 2017_slavin} to the regulation of star formation and galaxy evolution \citep[see, e.g.,][for reviews]{mckee07, naab17}.

In the classical picture, a SN remnant evolves through four main stages: free expansion, the Sedov--Taylor (ST) phase, the pressure-driven snowplow phase, and the momentum-conserving snowplow phase (e.g., \citealt{cox_cooling_1972, woltjer_supernova_1972, chevalier_self-similar_1982, cioffi_dynamics_1988, ostriker_astrophysical_1988, truelove_evolution_1999}). After the swept-up ambient mass becomes comparable to the ejecta mass, the remnant enters the ST phase, during which the initial explosion energy is converted primarily into thermal energy behind the shock, while the shock radius evolves as $R_{\rm SN} \propto (E/\rho)^{1/5} t^{2/5}$, where $E$ and $\rho$ are the explosion energy and ambient density, respectively \citep{1946_sedov, 1959_sedov, 1950_taylor}. This creates an extended region of hot, over-pressurized gas that expands into the surrounding ISM. Eventually, the remnant transitions into the snowplow phase, where energy losses reduce the pressure support and the expansion becomes increasingly momentum-driven.

The consequences of the ST phase extend beyond the evolution of individual remnants as it largely determines the efficiency with which supernovae couple their energy to the ISM. In particular, the ST phase governs two key aspects of SN feedback: the generation of momentum and the production of hot (i.e., $T > 2 \times 10^4$ K) gas \citep{1998_thornton, martizzi_15, kim_15}. Both are sensitive to the properties of the ambient medium, as the density and the efficiency of radiative cooling regulate the duration of the ST phase and, consequently, the amount of momentum injected and thermal energy retained by the surrounding gas. The hot bubbles produced during this evolution can persist for up to several million years in low-density environments, continuing to influence the ISM long after the initial explosion by evaporating cold atomic and molecular clouds, and producing ionizing X-ray radiation \citep{cox_cooling_1972, cox_74, mckee_77, 1998_thornton}.

The broader significance of this process was recognized by \citet{mckee_77}, who argued that the classical two-phase ISM model of \citet{field_69}, in which the ISM consists of hot gas at $T = 10^4$ K and cold gas at $T < 300$K, could not be maintained at the observed Galactic supernova rate ($\sim 10 ^{-13} \rm  pc^{-3} yr^{-1}$). Instead, they proposed a three-phase ISM consisting of a hot, low-density medium (HIM), cold, neutral, relatively dense clouds (CNM), and a mixture of warm ionized medium (WIM) and warm neutral medium (WNM). In this picture, supernova remnants are responsible for producing and maintaining the hot phase, which occupies a substantial fraction of the ISM volume. This framework established supernovae as a fundamental driver of ISM structure as their feedback injects momentum and thermal energy, drives turbulence, disperses dense cold gas, and helps regulate star formation. Within this picture, \citet{mckee_77} predicted that the evaporation rate of cold clouds embedded within SN-heated gas depends strongly on the ambient conditions, with the evaporated mass scaling with the background hydrogen density as $M_{\rm ev} \propto \nh^{-4/5}$ (see their Equation 8), highlighting the importance of the surrounding ISM properties in regulating the production of hot gas. 

While this and related analytical models provide valuable physical insight, they necessarily rely on idealized assumptions, including simplistic ambient medium and approximate descriptions of radiative cooling. In contrast, the real ISM is highly turbulent and multiphase, exhibiting large density contrasts and complex, temperature-dependent cooling that can substantially alter the evolution of supernova remnants and the resulting phase transitions. Quantifying these effects therefore requires numerical simulations capable of following the coupled evolution of shock propagation, radiative cooling, and the multiphase ISM.

Previous numerical studies have primarily focused on global aspects of SN feedback, such as the total swept-up mass, the momentum injected into the ISM, or the maximum hot-gas mass, which have proven highly effective for characterizing the overall dynamical impact of supernova remnants. As the SN blast wave expands, shock heating produces gas with temperatures exceeding $T_{\rm hot} \sim 2\times10^4$ K, causing the hot-gas mass to grow with time. Eventually, radiative cooling becomes efficient, and the hot-gas mass begins to decline. The maximum hot-gas mass, $\mmh$, therefore corresponds to the peak value attained during this evolution. \citet{kim_15} demonstrated that this maximum mass of hot gas produced by a single SN is only weakly sensitive to environmental inhomogeneity, while still depending on global properties such as ambient hydrogen density and magnetic fields. They found the scaling $\mmh \propto \nh^{-0.29}$ for a uniform medium and $\mmh \propto \nh^{-0.33}$ for a two-phase medium.

Similarly, \citet{steinwandel_20} investigated the effects of metallicity, numerical resolution, and thermal conduction on SNR evolution in uniform-density environments. They also found that the maximum hot-phase mass scales with hydrogen ambient density as $\mmh \propto \nh^{-0.28}$ to $\nh^{-0.29}$, with the normalization depending on metallicity (i.e., sensitive to cooling). Consequently, remnants evolving in low-density environments can produce more than an order of magnitude more hot gas than those in dense media (see their Fig. 16). They further showed that thermal conduction reduces the hot-gas mass by up to $\sim40\%$, while having a comparatively minor impact on the chemical composition and momentum budget of the remnant.

More recently, \citet{guo2024} studied SN remnants evolving in cloudy multiphase media up until $t = 3 \times 10^4$ yr, with an average total ambient density $\langle n \rangle = 10\,\mathrm{cm^{-3}}$, including thermal conduction and radiative cooling. Contrary to \citet{kim_15}, they found that the maximum hot-gas mass depends strongly on whether the background is uniform or turbulent, with values of $\sim 200\,M_\odot$ in cloudy media compared to $\sim 800\,M_\odot$ in uniform environments. They attribute this factor of $\sim4$ difference to significant energy losses through radiative cooling at the shock--cloud interfaces in the turbulent, cloudy medium, which reduces the energy available to heat the surrounding gas. This highlights the importance of the structure of the ambient medium: although the total injected SN energy is the same, interactions with dense cloud structures can lead to substantial radiative losses and consequently a much smaller hot-gas reservoir.

Overall, these studies have shown that the amount of hot gas produced by an individual SN can vary substantially with the properties of the surrounding medium -- for comparable ambient densities, substantially different amounts of hot gas have been reported, particularly when comparing homogeneous and cloudy media. Moreover, while the density dependence of the maximum hot-phase mass appears relatively consistent across studies, its normalization can vary substantially, reflecting potential differences in the efficiency of radiative cooling. This motivates a systematic investigation of the physical conditions that regulate the production of hot gas. In particular, we seek to determine how the maximum hot-phase mass depends on the ambient density and phase structure, and how these dependencies change when the efficiency of radiative cooling is varied.

A complementary aspect that has received less attention are SN-driven phase transitions, i.e., the origin of the hot gas within the full thermodynamic evolution of the ISM. Rather than focusing only on direct transitions into the hot phase, this perspective follows the trajectories of gas through density–temperature phase space as supernova shocks compress, heat, and pressurize the surrounding interstellar medium, driving exchanges of mass between the cold, warm, and hot phases. While shock heating transfers material from colder phases into warmer and hotter gas, radiative cooling simultaneously enables gas to return to lower temperatures. In turbulent environments, phase transitions are further promoted by turbulent mixing, which continuously redistributes gas between the phases \citep[e.g.,][]{braun_12_model, Banerjee_2014, 2020_fielding, kocjan_rhythm_2026}. However, the extent to which the hot phase is assembled from initially cold, warm, or already hot material remains poorly understood. 

Quantifying the origin of the hot phase therefore provides a second, complementary motivation for this work, as it is particularly relevant for the development of physically motivated subgrid models of phase exchange in galaxy simulations, where evaporation between multiple ISM phases is often represented using simplified prescriptions. For example, the widely used model of \citealt{springel_03} (hereafter \sh), adopted in simulations such as IllustrisTNG \citep{springel_18}, MillenniumTNG \citep{Pakmor_2023}, AURIGA \citep{grand_17}, THESAN \citep{kannan_22_thesan} and others, treats the ISM as a two-phase medium consisting of cold clouds embedded in a hot ambient gas. The two phases are in a feedback-regulated pressure equilibrium, where shock-heating caused by SNe and gas cooling determine both the effective pressure and the star formation rate (SFR). In this model, the SN-driven cold-to-hot mass exchange is parameterized as a density-dependent efficiency that scales as $\propto \langle n \rangle ^{-4/5}$, inspired by the results of \citet{mckee_77}. The \sh model further assumes that this process is primarily driven by thermal conduction, with a normalization that is treated as a free parameter in modern implementations such as in the \arepo code \citep{springel_10, weinberger_20_arepo}.

While this approach has been highly influential, it is primarily based on idealized two-phase description of the ISM. In a realistic, multiphase medium, however, warm and hot gas can originate from shock-heating of both the cold and warm phases, while turbulence continuously redistributes material between them. This raises the question of how evaporation should be quantified in a multi-phase ISM, i.e., whether a single evaporation efficiency is sufficient to describe the resulting mass exchange or any complex dependence on the detailed phase structure of the ISM. A more detailed understanding of the phase origin of hot gas can therefore provide additional constraints for improving such sub-grid models and linking them more directly to the underlying physics of supernova-driven mixing and cooling.

Motivated by these two complementary questions, i.e., what physical conditions regulate the production of hot gas in a multiphase ISM, and how this process can be represented in subgrid models, we revisit the concept of SN-driven gas evaporation in a more general framework that explicitly accounts for the role of radiative cooling, in addition to resolving multiple ISM phases and tracking their contributions to the hot gas. Using a suite of 3D supernova explosion simulations spanning a wide range of background densities and ISM conditions, we quantify the maximum hot-phase ($T > 2\times10^4$ K) mass, $\mmh$, and decompose it according to its initial phase origins using Lagrangian tracer particles. By systematically varying the adopted cooling prescriptions, we quantify how radiative cooling regulates hot-gas production and relate $\mmh$ to the properties of the ambient medium, including the mean density, the initial phase structure, and the temperature-dependent cooling time. In doing so, we extend Sedov--Taylor-inspired fitting functions by explicitly incorporating the effects of radiative cooling.

The structure of the paper is as follows. In Section \ref{sec:meth}, we describe our methodology as well as the numerical setup used in this work. We present our results in Section \ref{sec:res}, while Sections \ref{sec:dis} and \ref{sec:con} contain the discussion and summary of our results, respectively. 

\section{Method} \label{sec:meth}

\subsection{Supernovae simulations setup} \label{sec:sim_setup}

\begin{figure*}
    \centering
    \includegraphics[trim=0mm 5mm 0mm 0mm, clip, width=0.97\textwidth]{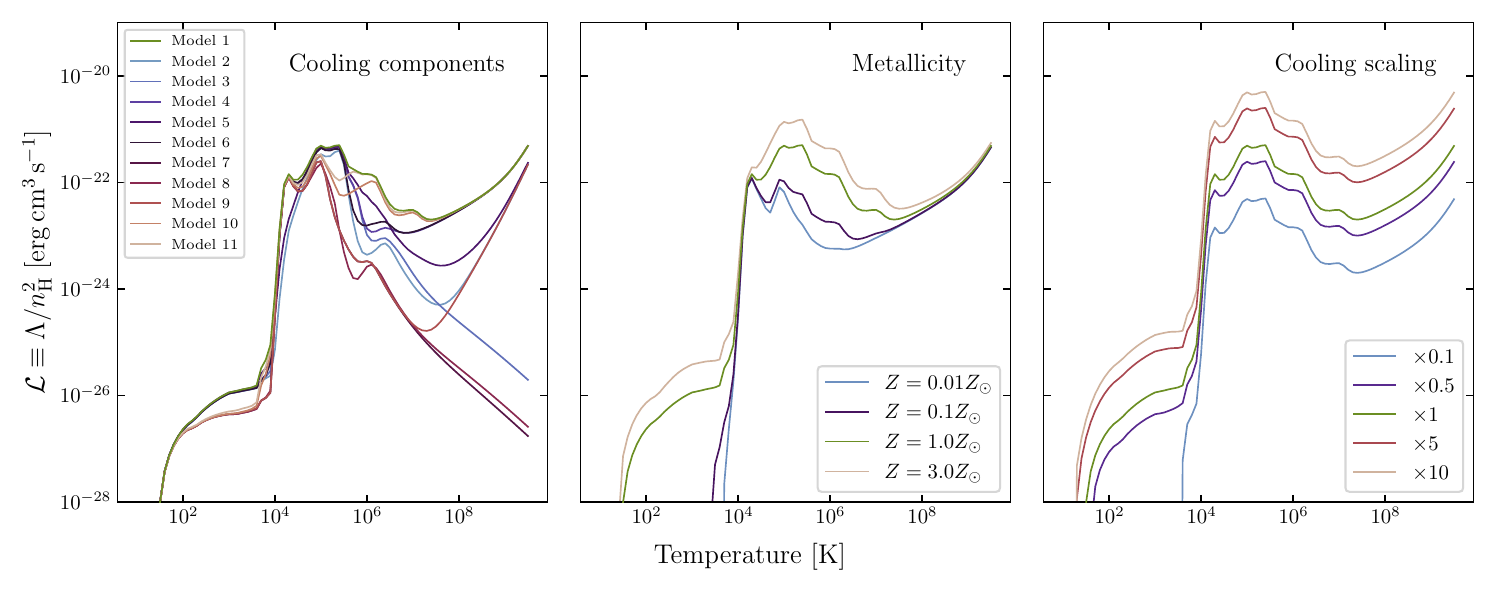}
    \caption{Cooling models considered in this work as a function of gas temperature, shown for $n_{\rm H}=1\,{\rm cm^{-3}}$ and $Z=1\,Z_{\odot}$ unless otherwise stated. We adopt the tabulated radiative cooling framework of \citet{ploeckinger_20}, which provides self-consistent cooling rates over a broad range of physical conditions. Our fiducial cooling model (Model 1; green line in all panels) corresponds to the full cooling prescription, including all available cooling processes, at solar metallicity $Z=1\,Z_{\odot}$. We explore three classes of modifications to this fiducial model: variations in the included cooling processes (left panel; see Table \ref{tab:cooling_models} in the Appendix), changes in gas metallicity (middle panel), and a uniform rescaling of the fiducial cooling function by factors of 0.1, 0.5, 1, 5, and 10 at all temperatures (right panel).}
    \label{fig:cooling_models}
\end{figure*}

\begin{figure*}[t]
    \centering 
    \includegraphics[trim=0mm 0mm 135mm 0mm, clip, width=0.98\textwidth]{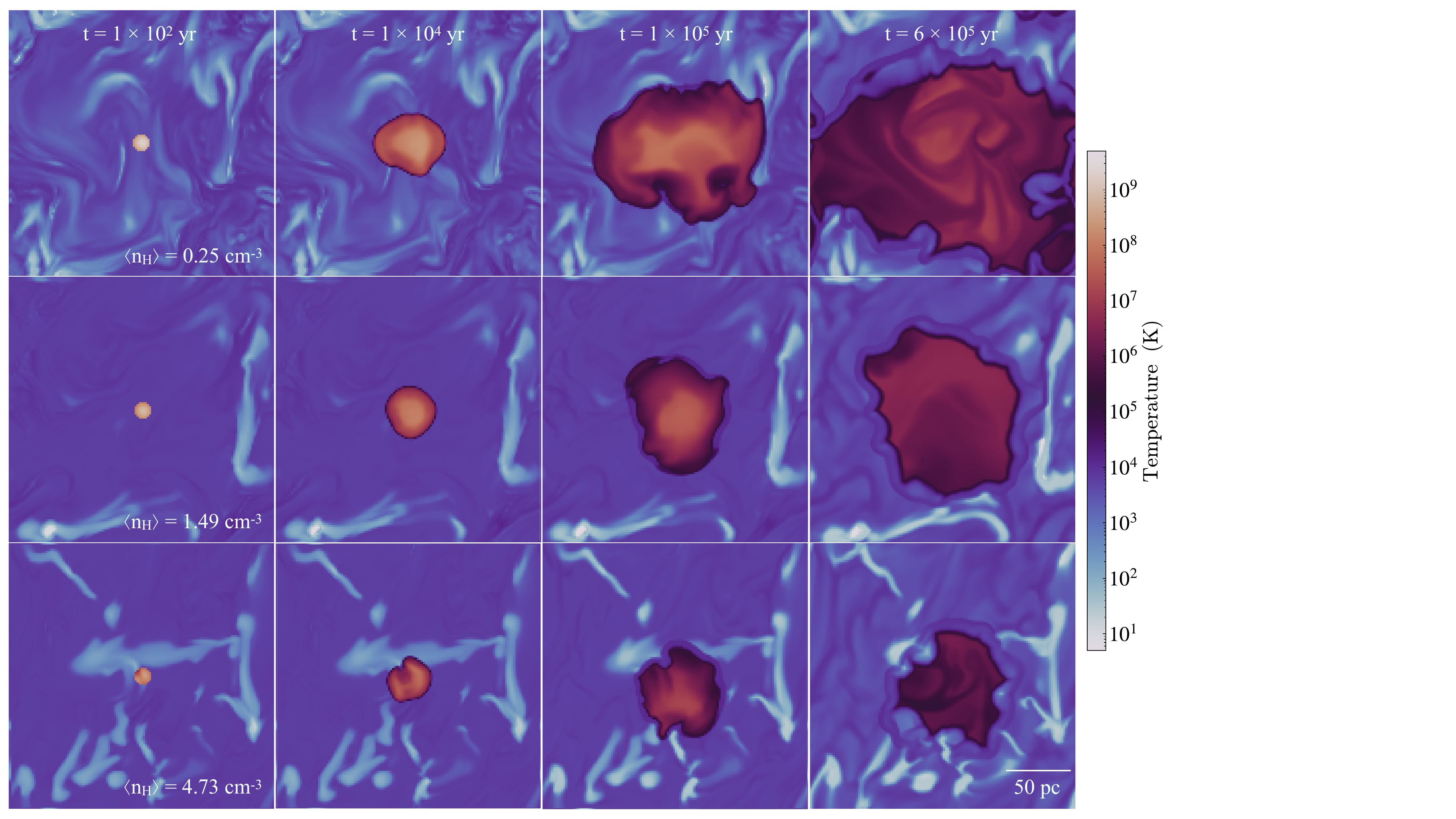}
    \caption{Gas temperature slice through the center of the box inside our turbulent-background SNe simulations, with box length of 200 pc and gas metallicity $Z\,= 1\,Z_{\odot}$. Each row corresponds to a different mean background density, while each column shows the shock-heated bubble at a different time, as labeled in the panels. As the SNR expands into the clumpy multiphase ISM, the bubble loses its spherical shape because of turbulent mixing between the shock front and the ambient gas, leading to cooling of the shock-heated material. Higher average background density results in stronger suppression of the SN bubble (see also Figure \ref{fig:fractions_evolution}).}
     \label{fig:combined_sn}
\end{figure*}

\begin{figure*}[t]
    \centering 
    \includegraphics[trim=8mm 168mm 30mm 50mm, clip, width=0.98\textwidth]{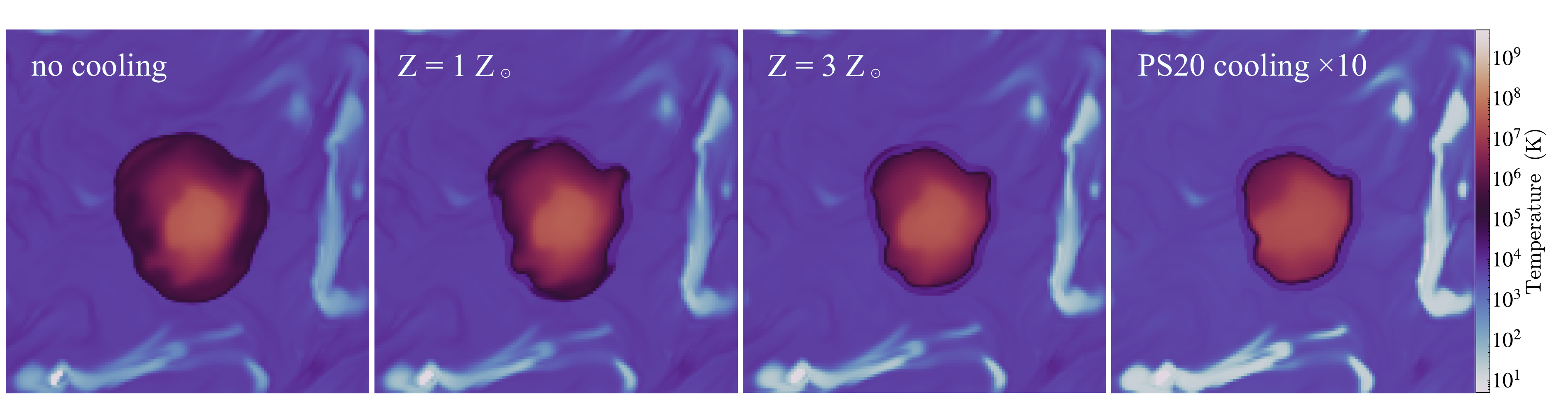}
    \caption{Same as Figure \ref{fig:combined_sn} but showing the SN remnant at a fixed time, $t=10^{5}\,{\rm yr}$, for simulations with $\nh =1.49\,{\rm cm^{-3}}$ and different cooling prescriptions. From left to right, the panels show a simulation without radiative cooling, the fiducial cooling model with $Z=1\,Z_{\odot}$ (see Model 1 in Table \ref{tab:cooling_models}), the same model but with $Z=3\,Z_{\odot}$, and a cooling function enhanced by a factor of ten (see right panel of Figure \ref{fig:cooling_models}). Despite the large variation in cooling efficiency, the overall size and morphology of the remnant remain broadly similar because the evolution is in the Sedov--Taylor phase, in which radiative losses have only a limited dynamical impact. The principal effect of stronger cooling is therefore a reduction in the thermal energy and hot-gas content of the remnant (see Figure \ref{fig:fractions_evolution}), resulting in a smaller hot bubble and a more pronounced shell. Conversely, in the absence of radiative cooling, the remnant retains more thermal energy, producing a larger hot interior and delaying the formation of the shell.}
    \label{fig:combined_sn_cooling}
\end{figure*}

\begin{figure*}[t]
    \centering 
    \includegraphics[width=0.98\textwidth]{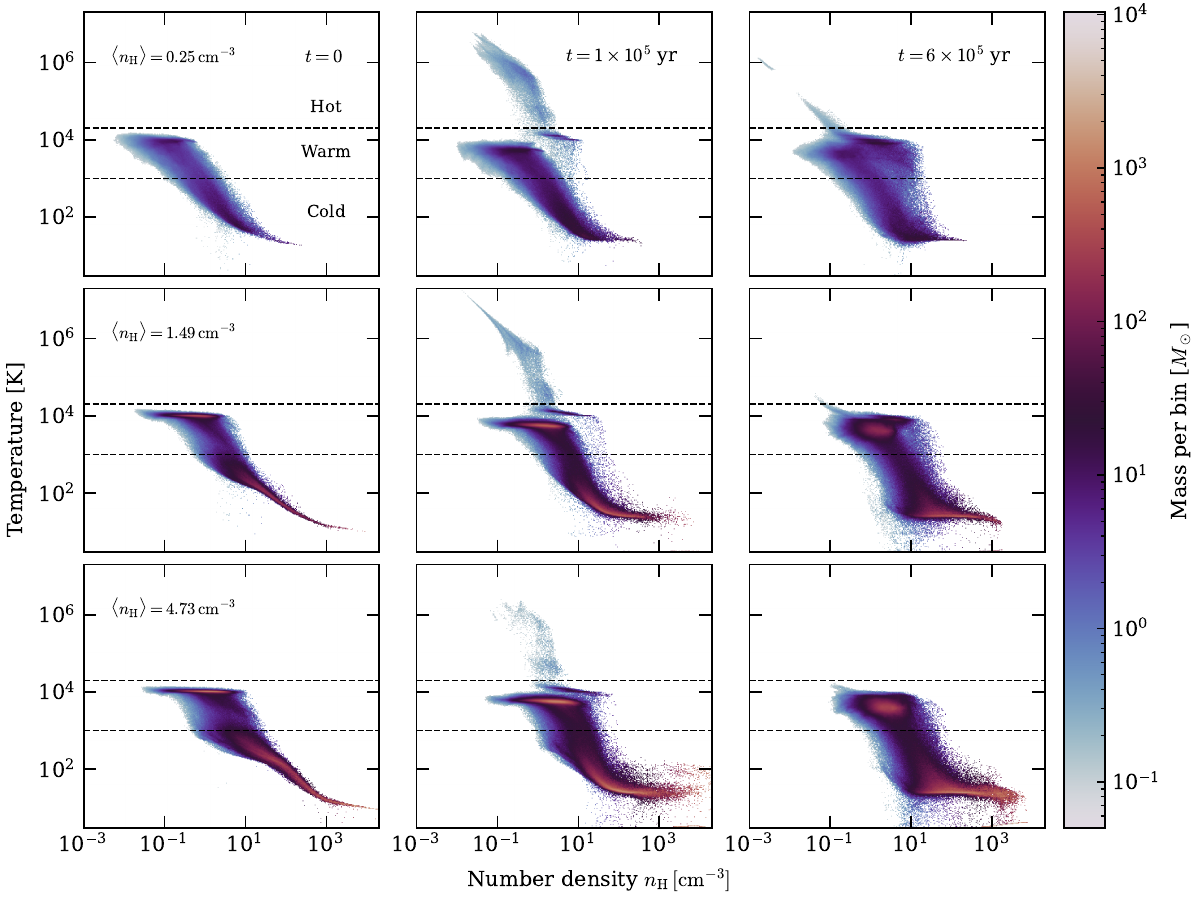}
    \caption{Temperature-density phase diagrams for the turbulent-background simulations. Each row corresponds to a different mean hydrogen number density, while each column shows a different time during the simulation, as labeled. The color scale indicates the gas mass contained in each $(n_{\rm H},T)$ bin. The upper dashed horizontal line marks the adopted lower limit of the hot-phase temperature, $T=2\times10^4$ K, while the lower dashed horizontal line marks the upper temperature limit of the cold phase, $T=10^3$ K. The warm phase occupies the intermediate temperature range between these thresholds. As the mean density increases, the gas distribution shifts toward lower temperatures, with an increasing fraction of the mass residing in the cold phase. The SN explosion heats the surrounding gas, causing it to enter the hot phase through evaporation. As the simulation progresses, this gas subsequently cools and transitions back into the warm and cold phases.}
     \label{fig:nT}
\end{figure*}

We use the adaptive mesh refinement (AMR) code \art (Adaptive Refinement Tree; \citealt{1997_kravtsov, 2008_rudd}) to simulate the evolution of a supernova explosion, from the Sedov–Taylor phase \citep{1946_sedov} through its subsequent radiative evolution. The simulations employ the default \art hydrodynamic solver, consisting of a directionally split Hancock scheme with second-order reconstruction, an exact Riemann solver, and the associated slope limiter (CG85; \citealt{1985CG}). All simulations are performed in three-dimensional boxes of width $L=200$ pc with a fixed grid resolution of $(2^7)^3 = 128^3$, corresponding to a cell size of $\Delta x_{\rm max} \approx 1.56\, \mathrm{pc}$, with no adaptive refinement. To ensure that our results are not affected by numerical resolution or by the adopted hydrodynamic solver, we additionally perform a subset of higher-resolution simulations and explore an alternative HLLC solver with linear reconstruction \citep{Leer_1983, toro_restoration_1994}, see Section \ref{sec:reso}. Although the problem could be treated in one dimension for uniform backgrounds, we adopt the same three-dimensional setup for all simulations to enable direct and consistent comparisons between idealized uniform media and realistic multiphase ISM environments.

We initialize the explosion with an injection of thermal energy ($10^{51}$ erg) at the center of the box, in a Gaussian kernel with characteristic radius $r_{\rm inj}=2\Delta x_{\rm max}$. The finite-sized Gaussian injection avoids grid-scale artifacts associated with depositing the energy into a single cell and produces a smooth initial pressure distribution. Since the injection region is much smaller than the characteristic scales governing the subsequent evolution, the solution rapidly converges to the Sedov--Taylor solution. This initialization intentionally skips the brief free-expansion phase, directly placing the remnant in the regime where the swept-up ambient mass exceeds the ejecta mass and the evolution is well described by the Sedov--Taylor solution. For a subset of simulations, we additionally vary the explosion location within the box to quantify the resulting scatter and assess the robustness of our results to environmental inhomogeneities (see Section \ref{sec:reso}).

Each simulation is evolved for 1 Myr, allowing the remnant to progress beyond the Sedov--Taylor stage. The code is run with passive Monte Carlo tracer particles that follow the gas flow by stochastically sampling mass fluxes between cells \citep{genel-2013, semenov-2018}, thereby providing access to the Lagrangian history of the gas. In all simulations, we adopt a tracer particle mass resolution of $1\,M_\odot$; we have verified that our results are insensitive to the choice of tracer particle mass over the range of values considered ($0.1-10\,M_\odot$). 

To model the radiative cooling of gas in our simulations, we adopt the tabulated framework developed by \citet{ploeckinger_20}\footnote{The base cooling setup corresponds to the UVB\_dust1\_CR1\_G1\_shield1 table from \url{https://www.sylviaploeckinger.com/radcool}, assuming redshift $z = 0$.}.  The tables provide the net cooling coefficient, $\mathcal{L}(T,Z) \equiv \Lambda(n_{\rm H}, T, Z)/n_{\rm H}^2$, where $\Lambda$ is the net radiative cooling rate per unit volume, over wide ranges of temperature, density, metallicity, and radiation fields. Since a wide range of cooling models are used in the literature, each including different subsets of radiative processes, it is challenging to explore the full space of possible cooling prescriptions. To this end, we construct a diverse suite of cooling models, as illustrated in Figure \ref{fig:cooling_models}. While several correspond to physically motivated cooling prescriptions, others are deliberately unphysical, allowing us to isolate the role of individual radiative processes in shaping the evolution of the blast wave and span a broader parameter space. Our fiducial cooling model is the full cooling prescription, including all available cooling processes (Model 1 in Table \ref{tab:cooling_models} and the left panel of Figure \ref{fig:cooling_models}; green line in all panels), evaluated at solar metallicity. We then construct a series of modified cooling models by varying one aspect of this fiducial prescription at a time. Specifically, we \emph{(a)} vary the included cooling processes (left column; see Table \ref{tab:cooling_models} in the Appendix for a list of included components in each model), \emph{(b)} vary the gas metallicity while retaining the full cooling prescription (middle column), and \emph{(c)} uniformly rescale the fiducial cooling function by factors of 0.1, 0.5, 1, 5, and 10 at all temperatures (right column). Together, these modifications produce a broad set of cooling functions that differ both in their overall normalization and in their temperature dependence, enabling us to robustly test how radiative loses influence supernova remnant evolution across the full temperature range relevant to the multiphase ISM. 

\subsection{Turbulent and uniform background modeling}

The initial conditions for the background ISM, namely the density, temperature, and velocity fields into which the SN explodes and propagates, are taken from the endpoints of multiphase ISM simulation runs with the AMR magnetohydrodynamic (MHD) code \textsc{RAMSES} \citep{teyssier_02}, with a setup that follows the works of \citet{bellomi_20}, \citet{godard_23} and \citet{MalamudEtAl-2026}. This particular setup simulates magnetized and partially ionized gas inside a 200 pc box. It was chosen given its success at reproducing a broad set of Galactic observables: \emph{(a)} the observed H and H$_2$ column density distributions; \emph{(b)} the probability distribution function of thermal pressure inferred from fine-structure excitation of carbon in the CNM; \emph{(c)} the velocity dispersion deduced from H\,{\sc i} emission spectra at high Galactic latitude \citep{godard_23}; \emph{(d)} the observed statistical abundance of CH$^+$ and its line profile distribution; and (e) the observed distribution of $\mathrm{OH^+}$, $\mathrm{H_2O^+}$ and H$_3^+$ \citep{MalamudEtAl-2026}.

The aforementioned observables were reproduced using the fiducial mean density of $\nh =1.49~\mathrm{cm^{-3}}$. Building on this established and validated baseline, we vary the mean density and UV field intensity (scaled by the parameter $G_0$ in \citet{MathisEtAl-1983} units), while keeping all other model parameters unchanged. Briefly, the setup is as follows (for more details see \citealt{godard_23}): the total ionization rate per H atom, including secondary ionizations of H by cosmic ray particles, is constant and equals $\zeta_\mathrm{H}=2\times10^{-16}\,\mathrm{s^{-1}}$. A homogeneous initial magnetic field $B$ is included (note that in this study $B$ is neglected and zeroed prior to the SNe simulations). Mechanical energy is injected in the gas at large scale. The amplitude of this turbulent forcing and the relative power injected in compressive modes are controlled by the forcing strength $F=1.5\times10^{-3}$ kpc\,Myr$^{-2}$ and compressive ratio $\xi=0.1$, as defined in \citet{godard_23}, resulting in a velocity dispersion $\sigma_v \sim 8$ km s$^{-1}$. 

Heating is induced by the photo-electric effect, cosmic ray particles, the formation of H$_2$ and its photo-destruction. Cooling is induced by the Lyman-$\alpha$ line, the fine structure lines of OI and CII, the recombination of electrons onto grains and the radiative cooling induced by the collisional excitation of H$_2$ rovibrational levels. Time-dependent chemical evolution considers: \emph{(a)} the formation of H$_2$ onto grains and its photo-destruction by UV photons, accounting for H$_2$ self-shielding and dust attenuation using the method of \citet{ValdiviaEtAl-2016}; \emph{(b)} H$^+$ formation by ionization of H, and recombination on PAHs, accounting for dust attenuation. PAHs ionization states are determined by photo-detachment, photoionization and by recombination with free electrons.

In this setup, changing $\langle n_{\rm H} \rangle$ alters the balance between photoelectric heating and radiative cooling, and thus the phase structure of the gas. To retain a two-phase (WNM and CNM) medium in all simulations, we therefore rescale the FUV intensity $I_{\rm UV}$ together with the density. The required scaling, $\nh \propto I_{\rm UV}^{1/2}$, deviates from the linear relation expected from simple heating-cooling balance because the grain and PAH charge, which sets the photoelectric heating efficiency, itself depends on $I_{\rm UV}$ \citep[][see also \citealt{shelest2026}, their Extended Fig.~1 and Eqs.~3--4]{wolfire_03}. Accordingly, our simulations span $\langle n_{\rm H} \rangle \approx 0.25$--$4.73\,\mathrm{cm^{-3}}$, corresponding to $I_{\rm UV} = 0.03$--$10$ (in Milky Way units), with each combination yielding a different initial cold-gas fraction, $f_{\rm cold,\,init}$.

To assess the impact of background inhomogeneities on SN-driven phase transitions, we perform a complementary set of simulations with a uniform medium, i.e., without any initial perturbations in gas density, velocity or temperature. The uniform simulations are initialized with a uniform background temperature of $T = 8 \times 10^3$ K. In the multiphase simulations, the temperature varies substantially with position across the different gas phases, but the volume-weighted average temperature of the box is also $\langle T \rangle \sim 8 \times 10^3$ K, across all density/UV regimes. We have tested the sensitivity of our results to this assumed initial temperature and found that the resulting hot gas mass is largely unchanged over the explored temperature range (see Section \ref{sec:max_mhot} for further discussion). Furthermore, we consider mean hydrogen densities in the range $0.1 \lesssim \nh \lesssim 5\,\mathrm{cm^{-3}}$, where numerical effects do not significantly influence the evolution of the SN remnant. At higher densities (e.g., $\sim 10\, \rm cm^{-3}$), the ambient thermal pressure associated with our adopted initial temperature strongly suppresses the SN bubble, leading to a reduced amount of hot gas produced, while at lower densities the remnant becomes comparable to the simulation box size and begins to overlap. 

In summary, each of our simulations corresponds to a unique combination of the background modeling (i.e., turbulent or uniform), mean gas density, and cooling prescription. In Figure \ref{fig:combined_sn}, we show temperature projections of the gas in some of our turbulent-background simulations following a supernova explosion. Each row corresponds to a different simulation, as labeled, all with $Z = 1 Z_{\odot}$. As the simulation progresses, the bubble expands and its morphology becomes less spherical due to turbulent mixing. The initially hot, shock-heated gas cools efficiently and begins to mix with the ambient warm and cold gas, leading to the development of a complex, multiphase structure. In Figure \ref{fig:combined_sn_cooling}, we compare the evolution of SN remnants at fixed time with several of the cooling models explored in this work, as well as without cooling (first panel), i.e., all of the components listed in Table \ref{tab:cooling_models} turned off. The slices show the remnant at $t = 10^5$ yr, in simulations with background density $\nh = 1\,\mathrm{cm^{-3}}$. Although the cooling prescriptions span more than an order of magnitude in cooling efficiency, the overall size and morphology of the remnant remain similar. This is expected because, at this stage, the remnant is still in the Sedov-Taylor phase, during which radiative cooling has only a modest influence on the global dynamics. Instead, the primary effect of the cooling prescription is on the thermal evolution of the shocked gas: stronger cooling reduces the amount of hot gas and promotes an earlier formation of a dense cooling shell, whereas weaker cooling allows the hot interior to persist for longer.

In the following, we define the cold phase to include gas with temperatures below $T_{\rm{cold}} = 10^{3}$ K. Following \citet{kim_15}, we define the hot phase as gas with temperatures exceeding $T_{\rm hot}=2\times10^4$ K. The warm phase occupies the intermediate temperature range $T_{\rm{cold}} < T < T_{\rm{hot}}$. Figure \ref{fig:nT} illustrates the resulting temperature--density phase structure for the turbulent-background simulations. The three temperature intervals defined above broadly correspond to the classical cold neutral medium (CNM), warm neutral medium (WNM), and hot ionized medium (HIM), although the exact correspondence is not one-to-one because our phase definitions are based solely on temperature. As the mean background density increases, the gas distribution shifts toward lower temperatures and a progressively larger fraction of the mass resides in the cold phase. Following the SN explosion, gas is heated into the hot phase through shock heating and evaporation of cold material. As the remnant evolves and radiative cooling becomes increasingly efficient, this gas subsequently transitions back into the warm and cold phases. At late times, the cold and warm gas distributions broaden toward lower densities, reflecting the disruption and dispersal of dense structures by the expanding SN remnant.

\section{Results} \label{sec:res}

A commonly adopted approach to quantify the outcome of supernova-driven phase transitions in the literature is to measure the maximum mass of the hot phase after a single (or a series of) SN explosion, which is typically reached after a fraction of Myr from the explosion, depending on the background gas density. However, this quantity \emph{(a)} does not reveal which ISM phase the shock-heated gas originated from, \emph{(b)} depends on the exact definition of hot gas, and \emph{(c)} can underestimate the cumulative impact of feedback when radiative cooling is efficient and rapidly removes material from the hot phase. An alternative approach is to follow gas transitions explicitly, for example by tracking the temperature evolution of Lagrangian gas parcels, and to count every instance in which material is heated from colder to hotter phases as the SN shock-wave expands. This approach captures the full history of feedback-driven heating, but may overestimate its effective impact if cooling operates on timescales shorter than those resolved by the sub-grid model because it counts short-lived heating events, even when they do not contribute to a sustained hot phase. Therefore, these two perspectives provide complementary but systematically different quantification of stellar feedback -- the peak/maximum and the integrated/total effect. 

In this work, we measure both the maximum hot-phase mass and the integrated mass transfer, and show that the latter can be understood in terms of the former and the initial distribution of gas phases. We first focus on determining the maximum mass of the hot phase (Section \ref{sec:max_mhot}), before quantifying the efficiency of the SN-driven mass transfer (Section \ref{sec:efficiency}).

\subsection{Gas-phase evolution} \label{sec:phase_ev}

\begin{figure*}
    \centering
    \includegraphics[width=0.85\textwidth]{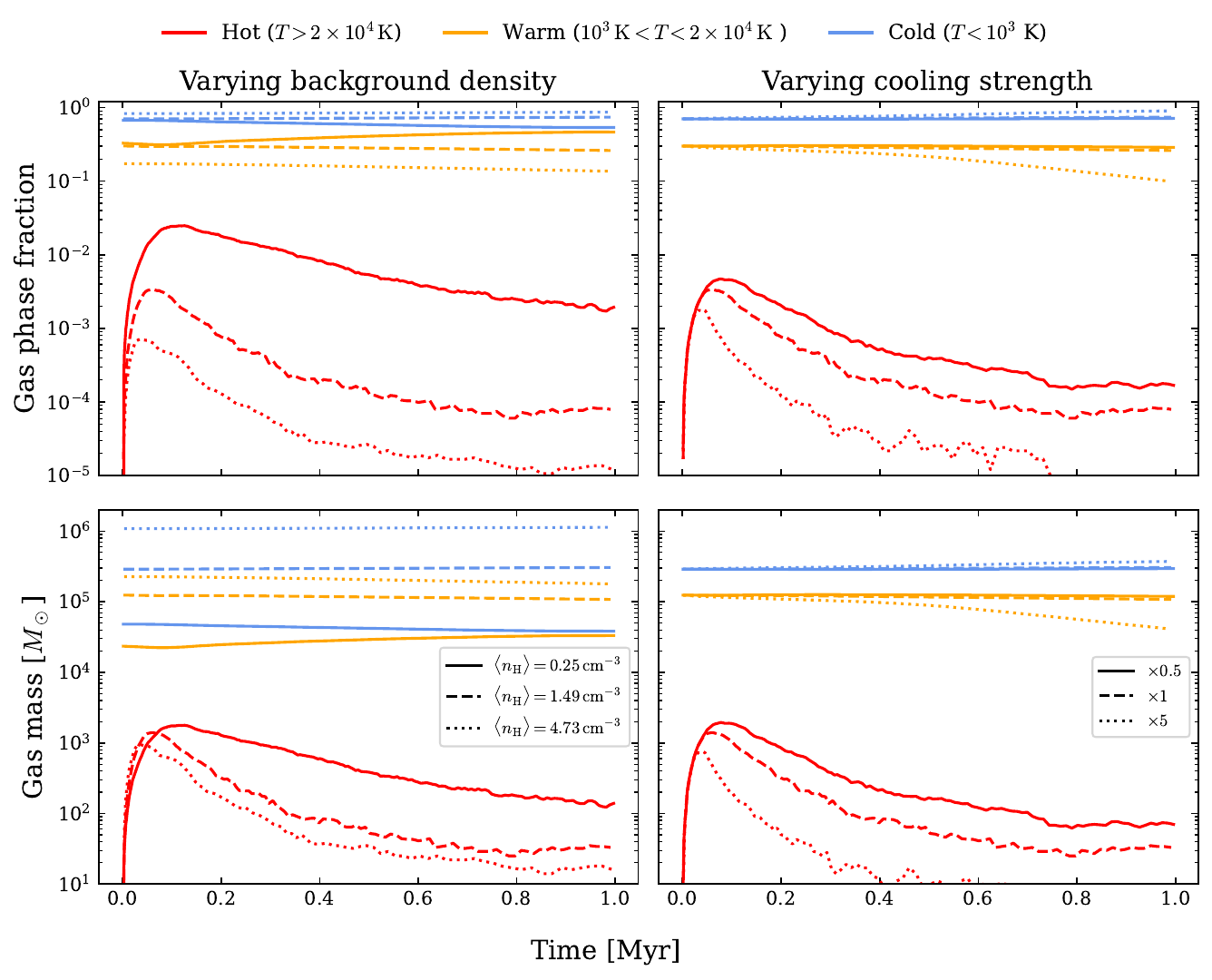}
    \caption{Evolution of the mass-weighted phase fractions (top row) and total phase masses (bottom row) for turbulent-ISM simulations. The hot, warm, and cold phases are shown in red, orange, and blue, respectively. The left column shows simulations with varying background density at fixed cooling (the fiducial model with $Z=1\,Z_{\odot}$), while the right column shows simulations with varying cooling strength at fixed density, $\nh=1.49\,{\rm cm^{-3}}$. Line styles indicate the average density (left column) or cooling-strength multiplier (right column), as labeled. In all cases, the hot phase grows rapidly following the SN explosion and reaches a maximum at $t\lesssim0.2\,{\rm Myr}$. Both increasing density and increasing cooling efficiency suppress the build-up of hot gas, reducing both the peak hot-phase fraction and the peak hot-gas mass. The warm and cold phases evolve correspondingly, reflecting the redistribution of mass driven by shock heating, radiative cooling, and turbulent mixing.}
    \label{fig:fractions_evolution} 
\end{figure*}

In Figure \ref{fig:fractions_evolution}, we show the evolution of the mass-weighted fractions (top panel) and total masses (bottom panel) of the hot (red), warm (orange), and cold (blue) phases for three representative turbulent-background SN simulations with average densities as labeled. For consistency with the subsequent analysis, these quantities are measured by tracking the thermal evolution of tracer particles throughout each simulation --- an equivalent analysis based on the Eulerian grid cells produces the same trends and quantitatively consistent phase fractions. The mass-weighted fractions are shown to illustrate the relative redistribution of gas among the different phases, while the total masses quantify the absolute amount of material contained in each phase and provide the more direct measure of the SN-driven phase evolution. We note that the phase fractions depend on the total gas mass in the simulation volume and therefore on the chosen box size. The left column compares simulations with different background densities and the fiducial cooling model, while the right column shows simulations with different cooling strengths at fixed density, $\nh = 1.49\, \rm cm^{-3}$. 

In all cases, the hot-phase fraction rises rapidly after the SN explosion and reaches a maximum at $t \lesssim 0.2\,{\rm Myr}$, in agreement with the literature \citep[e.g.,][]{kim_15, steinwandel_20, guo2024}. As expected, the amplitude of this peak depends on both the ambient density and the strength of radiative cooling, with the latter having a stronger influence --- the peak hot-gas mass exhibits only a moderate density dependence, varying by factors of a few across the explored parameter space. The differences become more pronounced at later times: while lower-density environments generally produce larger peak hot-gas masses, they sustain a much larger hot-gas reservoir throughout the subsequent evolution, whereas stronger cooling suppresses the growth of the hot phase and accelerates its decline. 

Following the peak, the hot gas gradually cools and transitions back into the warm and cold phases, with more efficient cooling leading to a faster transfer of material out of the hot phase. The evolution of the warm and cold phases reflects this redistribution of mass; in the lower-density and weaker-cooling simulations, the growth of the hot phase is accompanied by a steady decline in the cold-gas mass and a corresponding increase in the warm-gas reservoir, indicating efficient heating of the ambient medium by the expanding remnant. In contrast, simulations with stronger cooling or higher densities exhibit much smaller hot-gas masses and only modest changes in the warm and cold phases. These trends demonstrate that the amount of hot gas produced by a SN is regulated not only by the density of the surrounding medium, but also by the efficiency with which thermal energy is radiated away.

We note that cold--warm exchange might constitute a major component of the overall phase evolution. While our tracer analysis also allows us to quantify transitions between the cold and warm phases, we do not consider them further in this work. Unlike the production of hot gas, these transitions occur frequently in both directions and are strongly influenced by turbulent mixing and the local ISM environment, with many taking place independently of the SN explosion itself. As a result, it is difficult to isolate the direct impact of SN feedback on the cold--warm exchange. We therefore focus on the production of hot gas, where the connection to the SN is more direct, and defer a detailed analysis of cold--warm phase exchange to future work.

\subsection{Maximum mass of the hot phase} \label{sec:max_mhot}

In the following, we focus on the maximum mass of hot gas, $M_{\rm max,\, hot}$, which we define as the maximum mass of gas with $T > T_{\rm hot}$ reached over the time evolution. We determine it by identifying the simulation snapshot at which the hot-gas mass reaches its peak. We have verified that performing the same measurement using grid cells instead of tracer particles provides consistent results.

\subsubsection{Cooling effects on the maximum hot-phase mass} \label{sec:max_mhot_cool}

As described in Section \ref{sec:intro}, an intuitive expectation is that $M_{\rm max,\, hot}$, a quantity closely related to the maximum swept-up mass of the remnant, should decrease with increasing ambient density as denser gas cools more efficiently and is more difficult to evaporate. However, Figure \ref{fig:fractions_evolution} shows that $M_{\rm max,\,hot}$ cannot be described by a density dependence alone. While increasing the ambient density suppresses the hot phase, a similar suppression is obtained by strengthening the cooling function at fixed density. This indicates that the quantity is fundamentally controlled by radiative energy losses and correlating $\mmh$ solely with the mean background density would overlook the impact of variations in metallicity and the detailed shape of the cooling function. 

Motivated by this, we seek a more direct connection between $\mmh$ and the efficiency of radiative cooling. Since the cooling rate is a strong and highly non-linear function of temperature, it is more informative to consider its cumulative effect over the full thermal evolution of the gas than its instantaneous value. We therefore characterize the cooling efficiency using the density-normalized integrated isobaric cooling time $\tcool$, which measures the total time required for gas to cool from $T_{\rm max}$ to $T_{\rm min}$ at constant pressure. By using this definition, we factor out the density dependence introduced by the cooling rate, allowing the resulting quantity to characterize the cooling efficiency independently of it. To derive this quantity, we start from the temperature evolution equation for an ideal monatomic gas ($\gamma = 5/3$) undergoing isobaric radiative cooling \citep{draine_11},
\begin{equation}\label{eq:dTdt}
    \frac{dT}{dt} = \frac{\Lambda}{(5/2) n_{\rm H} k_{\rm b}}.
\end{equation}
Rearranging Equation \ref{eq:dTdt} gives
\begin{equation}\label{eq:dt}
dt = \frac{5}{2}\frac{n_{\rm H} k_{\rm B}}{\Lambda}dT
= \frac{5}{2}\frac{k_{\rm B}}{n_{\rm H}\mathcal{L}}dT,
\end{equation}
where we have used the volumetric cooling rate $\Lambda(n_{\rm H},T,Z)=n_{\rm H}^2\mathcal{L}(T,Z)$, where $\mathcal{L}$ is the cooling function (see Figure \ref{fig:cooling_models}). To obtain the cooling time, we now integrate the right-hand side over the temperature range of interest. We assume that the cooling is isobaric, such that the pressure remains constant, $P=P_0=n_0k_{\rm B}T_0$. Therefore, it follows that $n_{\rm H}T = n_{\rm H,0}T_0$, where $n_{\rm H,0}$ and $T_0$ are the initial hydrogen number density and temperature, respectively. Substituting this relation into Equation \ref{eq:dt} and integrating on both sides we find 
\begin{equation}\label{eq:tcool_def}
    t_{\rm cool} = \frac{5}{2}\frac{k_{\rm B}}{T_0\, n_{\rm H, 0}} \int_{T_{\rm min}}^{T_{\rm max}} \frac{T'}{\mathcal{L}(T')} \,\mathrm{d}T'  \,.
\end{equation}
Multiplying through by the initial number density $n_{\rm H,0}$ defines the approximately
density-independent cooling time,
\begin{equation}\label{eq:tcool_int}
\tcool \equiv t_{\rm cool} \, {n_{\rm H,0}} = \frac{5}{2}\frac{k_{\rm B}}{T_0} \int_{T_{\rm min}}^{T_{\rm max}} \frac{T'}{\mathcal{L}(T')} \,\mathrm{d}T'.
\end{equation}
In this work, we take the initial hydrogen number density to be the average hydrogen number density of the initial ISM-box conditions, $n_{\rm H,0} = \nh$, while the initial temperature is equal to the volume-weighted average temperature in our turbulent-background simulations $T_0 = \langle T \rangle \approx 8000\,\mathrm{K}$. This definition of $\tcool$ removes the explicit density normalization from the integrated cooling time; the remaining, weak dependence on density enters only through the cooling function itself. The quantity therefore provides a convenient summary of the cooling efficiency over the temperature interval $T_{\rm min}$--$T_{\rm max}$. It should not, however, be interpreted as the actual time required for a gas parcel in the simulations to cool through this temperature range. Equation \ref{eq:tcool_int} assumes idealized isobaric cooling, whereas gas in the simulations evolves dynamically, with both density and pressure changing in response to turbulence, shocks, and radiative losses. The integrated cooling time should therefore be regarded as a characteristic measure of the cumulative cooling efficiency over the selected temperature interval, rather than the true thermodynamic evolution time of individual gas parcels.

The appropriate integration limits are, however, not known a priori. During the evolution of a supernova remnant, the shocked gas spans a broad range of temperatures that changes continuously as the shock expands. Rather than adopting an arbitrary temperature interval, we determine empirically which range is most relevant for regulating $\mmh$. Specifically, we identify the values of $T_{\rm min}$ and $T_{\rm max}$ for which the integrated cooling time exhibits the strongest correlation with the maximum hot-gas mass.

\begin{figure*}
    \centering 
    \includegraphics[trim=12mm 38mm 10mm 25mm, clip, width=\textwidth]{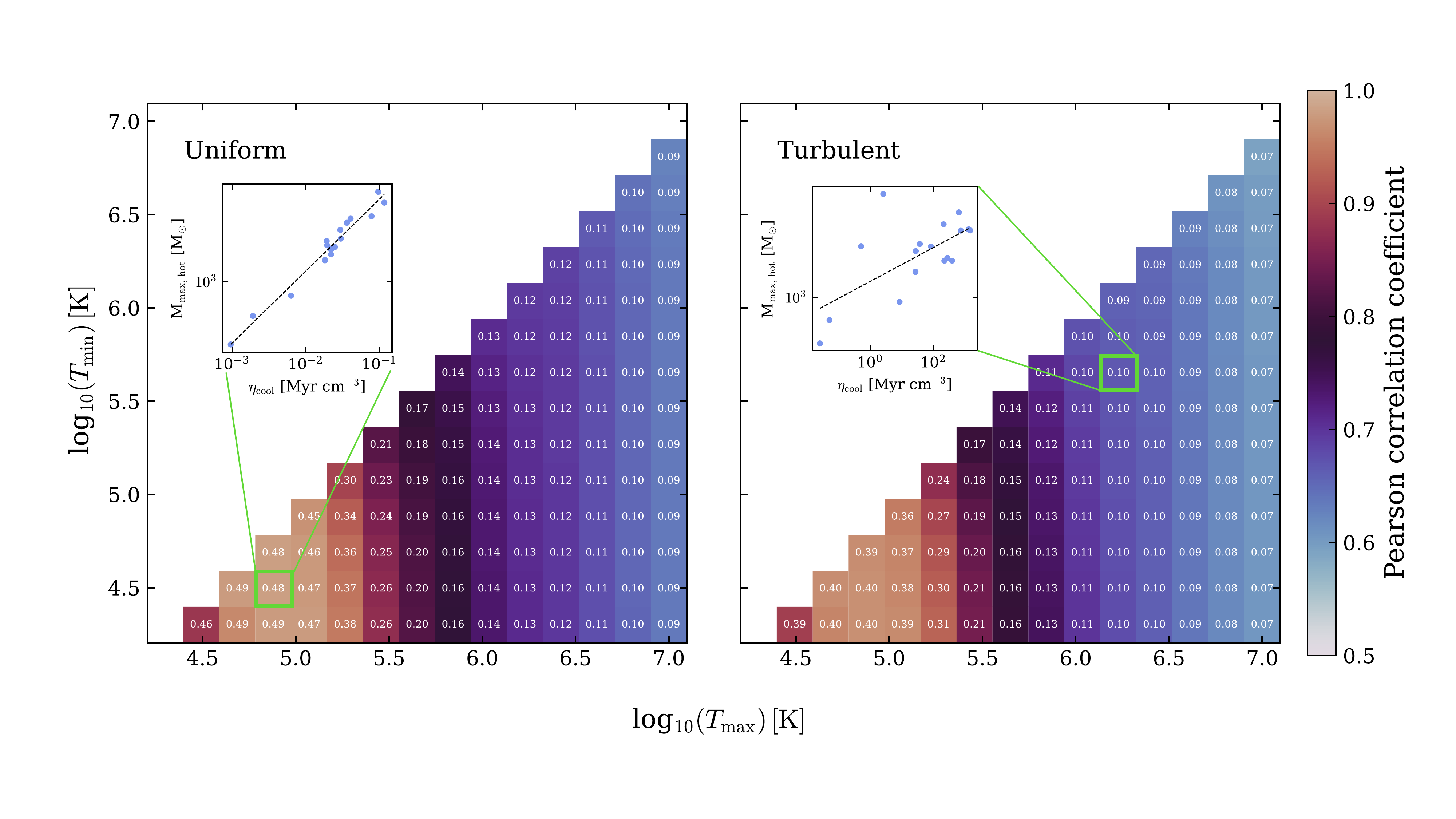}
    \caption{Correlation between $\mmh$ and the integrated cooling time $\tcool \equiv t_{\rm cool} \nh$ (see Equation \ref{eq:tcool_int}) over different temperature intervals $(T_{\rm min}, T_{\rm max})$. Each pixel corresponds to a particular choice of $T_{\rm min}$ (y-axis) and $T_{\rm max}$ (x-axis). For each temperature range, we fit a power-law relation between $\mmh$ and $\tcool$. The colormap shows the resulting Pearson correlation coefficient, while the overlaid number indicates the best-fit power-law slope. Example $\mmh$--$\tcool$ relations are shown in the upper-left corner of each panel. Results are shown for simulations with $\nh=1.49\,{\rm cm}^{-3}$, with a turbulent background (left) and a uniform background (right) and all the cooling models considered (Figure \ref{fig:cooling_models}). Regions with higher correlation coefficients identify the temperature ranges most relevant for determining $\mmh$. Across all simulations considered, we find the strongest correlation to be for $T_{\rm min}=10^{4.5}\,\mathrm{K}$ and $T_{\rm max}=10^{5.1}\,\mathrm{K}$.}
    \label{fig:triangles}
\end{figure*}

To this end, we first compute the Pearson correlation coefficient between $\mmh$ and the integrated cooling time evaluated over different temperature intervals. Specifically, for each pair of temperatures $(T_{\rm min}, T_{\rm max})$, we calculate the integrated cooling time within that range and fit a power-law relation to the resulting $\mmh$ values. In Figure \ref{fig:triangles}, each pixel corresponds to a particular choice of $T_{\rm min}$ ($y$-axis) and $T_{\rm max}$ ($x$-axis). The color indicates the Pearson correlation coefficient of the fit, while the overlaid value shows the corresponding power-law slope. The left and right panels show the results for the uniform and turbulent simulations, respectively, for a representative case with a mean hydrogen density of $\nh = 1.49\,\rm cm^{-3}$.
This analysis is repeated independently for every mean background density and for both the uniform and turbulent simulation suites. For each combination of background type and average density, we evaluate all of the cooling models shown in Figure \ref{fig:cooling_models}.

We find that, across all background types, mean densities and the cooling models considered, the strongest correlation is obtained for the temperature interval $T_{\rm min}=10^{4.5}\,\mathrm{K}$ and $T_{\rm max}=10^{5.1}\,\mathrm{K}$, as seen in the colors in the figure. The physical significance of this particular temperature interval is discussed below. We therefore adopt this range when constructing the correlation between the integrated cooling time $\tcool$ and $\mmh$, which we obtain by fitting the scaling relation $\mmh \propto \nh^{\alpha}\, {\tcool}^{\beta}$, where $\nh$ is the mean hydrogen background density of the initial conditions. We obtain two very similar scaling relations for the maximum hot-gas mass in the uniform and turbulent runs:
\begin{equation} \label{eq:mhot_max_uni}
M_{\rm max,\, hot,\, uni} = 1.08 \times 10^4\, M_{\odot}\, \left(\frac{\nh}{1\rm cm^{-3}}\right)^{-0.29}\,\left(\frac{\tcool}{1 \rm Myr\, cm^{-3}}\right)^{\,0.44}.
\end{equation}
\begin{equation}\label{eq:mhot_max_turb}
M_{\rm max,\, hot,\, turb} = 1.05 \times 10^4 \, M_{\odot}\, \left(\frac{\nh}{1\rm cm^{-3}}\right)^{-0.30} \,\left(\frac{\tcool}{1 \rm Myr\, cm^{-3}}\right)^{\,0.38}.
\end{equation}
These scalings are numerically derived from our simulations; in Section \ref{sec:analytical}, we complement this numerical result with an idealized analytic model that provides physical insight into why the maximum hot-phase mass depends on these two quantities and, in particular, why the resulting power-law indices take the values found in the simulations. 

The similarity of the resulting exponents in the uniform and turbulent simulations suggests that the maximum hot-gas mass is governed primarily by the ambient density and the integrated cooling time of gas in the $10^{4.5}$--$10^{5.1}\,\mathrm{K}$ temperature range, with only a weak dependence on the structure of the surrounding medium. The positive dependence on $\tcool$ is expected, since larger integrated cooling time corresponds to less efficient radiative cooling, allowing gas to remain hot for longer and thereby increasing the maximum hot-gas mass $\mmh$. It is important to note that $\tcool$ in these relations is not the physical cooling time itself, but the density-independent integrated cooling time defined in Equation \ref{eq:tcool_int}. By construction, the explicit dependence on the initial density is removed through the multiplication by $n_{\rm H,0}$. Therefore, unlike the usual cooling time $t_{\rm cool}$, which scales approximately as $t_{\rm cool}\propto \nh^{-1}$ for fixed temperature, $\tcool$ does not inherit this leading density dependence --- its remaining dependence on density arises only indirectly through the cooling function. The fitted scaling relations therefore imply that the maximum hot-gas mass has an intrinsically weak, shallow power-law dependence on ambient density: $M_{\rm max,hot}\propto \nh^{\alpha}$ with $\alpha\simeq -0.3$ for both the uniform and turbulent simulations. This is consistent with previous studies \citep[e.g.,][]{kim_15, steinwandel_20} and indicates a modest influence of the background density on the dynamical evolution of the gas (see Figure \ref{fig:fractions_evolution}).

\begin{figure*}
    \centering 
    \includegraphics[trim=10mm 0mm 10mm 0mm, clip, width=0.85\textwidth]{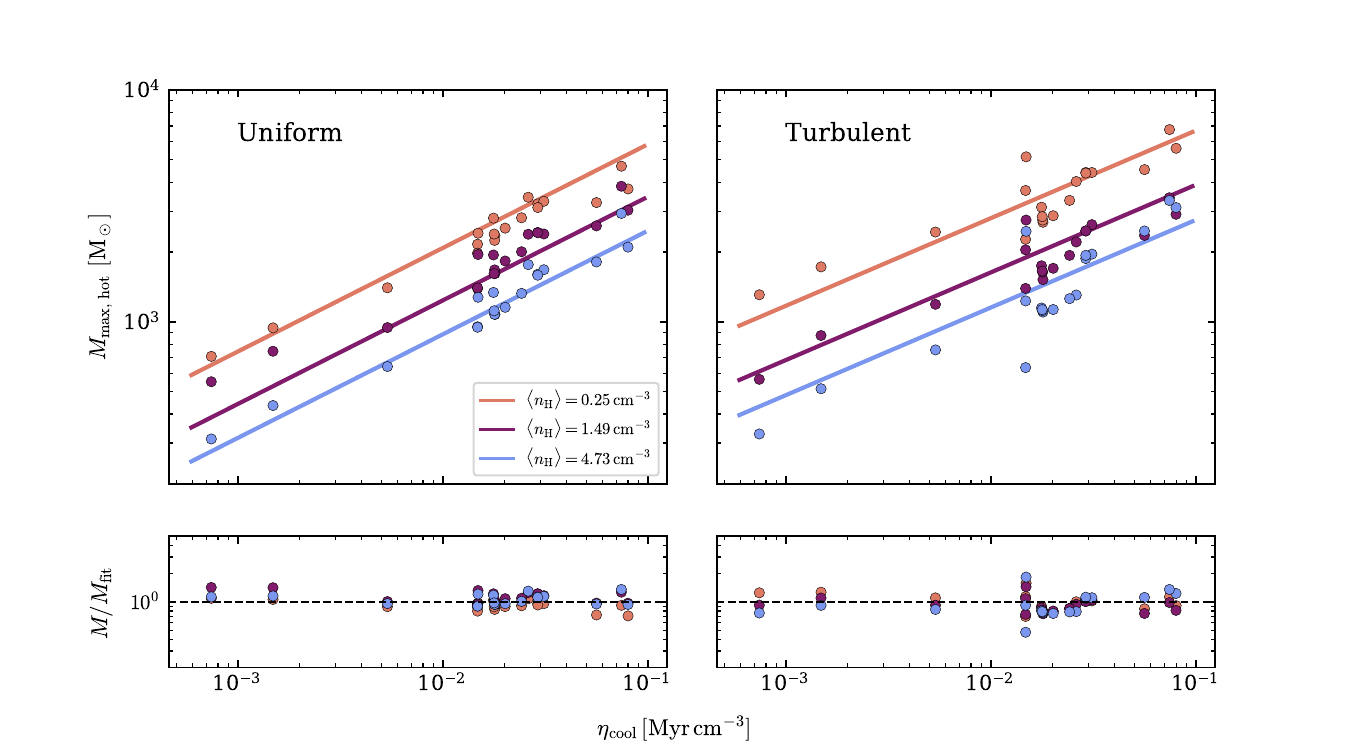}
    \caption{Maximum hot gas mass, $\mmh$, as a function of the integrated cooling time, $\tcool \equiv t_{\rm cool} \nh$ (see Equation \ref{eq:tcool_int}) The left panels show simulations with a uniform background medium, while the right panels show simulations with a turbulent background. Points represent the measured values of $\mmh$, and solid lines show the corresponding predictions from Equations \ref{eq:mhot_max_uni} (uniform) and \ref{eq:mhot_max_turb} (turbulent). The color of each point indicates the mean background hydrogen number density. The lower panels show the ratio of the measured to predicted values. The ratios remain close to unity over nearly two orders of magnitude in $\tcool$ and across the full density range considered, demonstrating that the fitting relations provide an accurate description of the simulation results with no evident systematic trends.}
    \label{fig:Mhot_tcool}
\end{figure*}

In Figure \ref{fig:Mhot_tcool}, we compare the derived scaling relations with the maximum hot gas mass, $\mmh$, measured in our simulations. The figure shows $\mmh$ as a function of the integrated cooling time, $\tcool$, for simulations spanning three background densities, indicated by the color scale. The left and right columns correspond to the uniform-background and turbulent simulations, respectively. The solid lines show the predictions of Equations \ref{eq:mhot_max_uni} and \ref{eq:mhot_max_turb}, while the points denote the measured values from the simulations.

The lower panels show the ratio between the measured and predicted values of $\mmh$ for each simulation. The ratios remain close to unity across the full range of $\tcool$ and background densities, with no obvious systematic dependence on either parameter. In Section \ref{sec:dis_lit}, we further place these results in the context of previous numerical studies (see Figure \ref{fig:summary_plot}).

\begin{figure}
    \centering
    \includegraphics[trim=0mm 5mm 0mm 0mm, clip,width=0.48\textwidth]{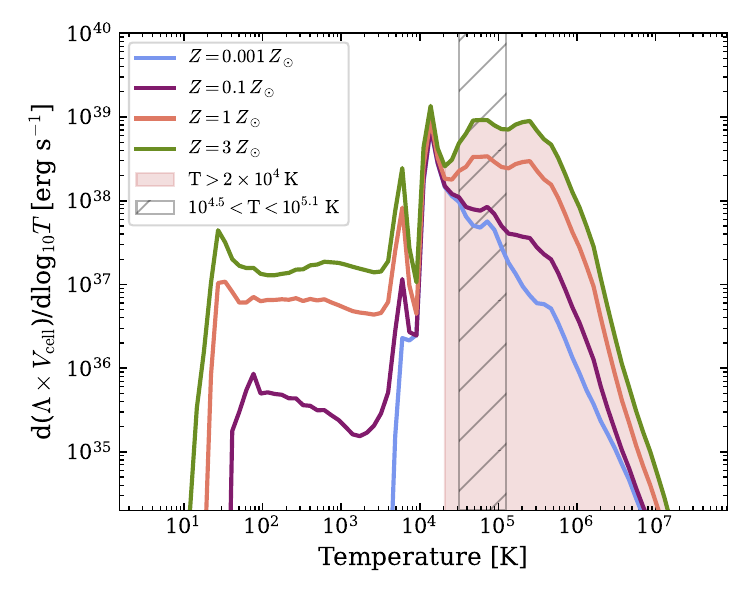}
    \caption{Radiative energy loss rate per logarithmic temperature interval, $d(\Lambda \times V_{\rm cell})/d\log_{10}T$, as a function of temperature for simulations with $\nh=1.49\,{\rm cm}^{-3}$ and metallicities indicated in the legend. The quantity is obtained by evaluating the cooling rate of each simulation cell and binning the resulting radiative energy losses over logarithmic temperature intervals. The dashed shaded region marks the temperature interval $(T_{\rm min}, T_{\rm max})$ which correlates the strongest with $\mmh$, while the red shaded region denotes the hot-phase temperature range. Although the overall cooling losses decrease with metallicity, the selected temperature interval consistently overlaps with the temperatures that contribute most strongly to the cooling of the hot gas.} 
    \label{fig:dEdtdT} 
\end{figure}

\subsubsection{The temperature range of dominant cooling losses} \label{sec:trange_dis}

To better understand why the temperature interval identified above correlates strongly with $\mmh$, we examine where radiative cooling is most effective across the relevant temperature range. Specifically, we show in Figure \ref{fig:dEdtdT} the distribution of radiative cooling losses as a function of temperature, ${\rm{d}}(\Lambda V_{\rm cell}){\rm{d}\,log_{10}}T$. This quantity is computed directly from the simulation outputs by evaluating the volumetric cooling rate in each cell, $\Lambda(n_{\rm H},T,Z)$, multiplying by the corresponding cell volume, $V_{\rm cell}$, and binning the resulting radiative energy losses over logarithmic temperature intervals. We show the results for the representative turbulent-background, $\nh = 1.49\,\rm cm^{-3}$ simulations with the cooling models with varied metallicity (middle panel in Figure \ref{fig:cooling_models}). 

The dashed region marks the temperature interval $(T_{\rm min}, T_{\rm max})$ that produces the strongest correlation with $\mmh$, while the red shaded region indicates the temperature range used to define the hot phase (i.e., $T > 2\times 10^4$ K). As expected, the overall cooling rate decreases with decreasing metallicity. Despite this variation, the selected $(T_{\rm min}, T_{\rm max})$ interval generally lies close to the temperatures where the bulk of the radiative energy losses occur. This result is not a consequence of the adopted hot-phase temperature threshold, but instead reflects the temperature range over which cooling most effectively reduces the energy available to drive the expansion of the supernova remnant --- the greater the amount of energy lost through radiation, the less energetic the shock becomes. Consequently, the bubble stalls at a smaller radius (see Figure \ref{fig:combined_sn_cooling}), thereby limiting the amount of ambient gas that is swept up and transferred to hotter phases (see Figire \ref{fig:fractions_evolution}). The cooling therefore affects $\mmh$ indirectly through the dynamics of the expanding remnant, in addition to directly transferring gas out of the hot phase.

\subsection{SN-driven gas evaporation efficiency} \label{sec:efficiency}

We now focus on quantifying the efficiency of SN-driven mass evaporation. In large-volume galaxy formation simulations, SN-driven unresolved mass exchange between ISM phases is commonly described through analytical prescriptions, where the amount of gas transferred by a single feedback event is parameterized rather than computed directly, as this process occurs below the resolution scale. This is typically expressed in terms of an evaporation efficiency, which measures how much gas is transferred from one phase to another per feedback-emitting star. Such a quantity provides a dimensionless characterization of the strength of SN-driven mass exchange. A widely used example is the evaporation efficiency parameter introduced by \citet{springel_03}. In the \sh model, the ISM is represented as a two-phase medium consisting of cold clouds embedded within a hot ambient phase, so naturally only cold-to-hot evaporation is considered. The corresponding evaporation efficiency factor $A_{\rm ch}$ is defined as
\begin{equation}\label{eq:mckee77}
    A_{\rm SH03,\ ch} = A_0\, \left( \frac{\rho}{\rho_{\rm{SF}}} \right)^{-0.8},
\end{equation}
following the scaling proposed by \citet{mckee_77}. This quantity further enters the cold-to-hot evaporation rate, which is proportional to the mass fraction $\beta$ of feedback-producing stars and the star-formation rate, $\propto 1/t_{\rm *}$, where $t_{\rm *}$ is the star-formation timescale (proportional to the local dynamical time of the gas). The quantity $A_0$ controls the overall evaporation strength and is treated as a free parameter in galaxy simulations \citep[e.g., $A_0 = 573$ in IllustrisTNG][]{vogelsberger_14_illustris, pillepich_18_tng, nelson_19_datarelease}, while $\rho_{\rm SF}$ is the star-formation density threshold, corresponding to approximately $n_{\rm H}\sim0.1\,{\rm cm^{-3}}$. This definition of evaporation efficiency assumes that the input density corresponds to that of dense, cold clouds and accounts only for evaporation by thermal conduction, neglecting the potential contribution from turbulence. Moreover, it is restricted to evaporation from the cold to the hot phase and therefore cannot describe the multiphase transitions captured in our simulations.

Therefore, to extend SN-driven mass exchange beyond the cold-to-hot evaporation, we define a generalized evaporation efficiency factor,
\begin{equation}
A_{\rm xy} = \frac{M_{\rm{xy}}}{\langle M_{\rm{fb\,star}} \rangle},
\end{equation}
where $M_{\rm xy}$ is the mass transferred from phase $x$ to phase $y$ and $\langle M_{\rm{fb\,star}} \rangle$ is the mean mass of stars that produce stellar feedback, i.e., stars massive enough to explode as core-collapse supernovae. Thus, $A_{\rm xy}$ represents the amount of gas transferred between phases per unit average feedback-emitting star mass, providing a direct measure of the efficiency of SN-driven phase exchange that is applicable to any pair of ISM phases. 
The mean feedback-emitting stellar mass is computed from the stellar initial mass function, $\phi(M)$, as
 \begin{equation}\label{eq:m_fb}
    \langle M_{\rm fb\,star} \rangle = \frac{\int_{8 M_{\odot}}^{M_{\rm max}} M \phi(M) \,dM }{\int_{8 M_{\odot}}^{M_{\rm max}} \phi(M) \,dM} \,.
\end{equation}
Assuming that such stars have masses above 8 $M_{\odot}$, while the maximum stellar mass is 150 $M_{\odot}$, the average feedback-star mass depends only on the high-mass slope of the adopted IMF. For a \citet{salpeter_55} IMF, where $\phi \propto M^{-2.35}$, $\langle M_{\rm fb\,star} \rangle$ is equal to 20.18 $M_{\odot}$. For the \citet{2001_kroupa} and \citet{2003_chabrier} IMFs, both of which have a high-mass slope $\phi \propto M^{-2.3}$, we obtain $\langle M_{\rm fb\,star} \rangle = 20.74\, M_{\odot}$. Throughout this work, we adopt the latter value.

In our simulations, the mass transferred between two ISM phases, $M_{\rm xy}$, can be defined in two complementary ways. 
The first one focuses on the peak hot-gas reservoir and considers only the mass that contributes to the hot phase at the time when the hot-gas mass reaches its maximum --- $M_{\rm max,\, xy}$. This defines the \emph{maximum} evaporation efficiency,
\begin{equation}\label{eq:A_max_xy}
A_{\rm max,\,xy} = \frac{M_{\rm max,\, xy}}{\langle M_{\rm fb,star}\rangle}.
\end{equation}
The second one takes into account the entire mass transferred between phases over the full simulation duration (1 Myr), which produces the \emph{total}, integrated evaporated mass $M_{\rm tot,\, xy}$, with the  efficiency 
\begin{equation}
A_{\rm tot,\,xy} = \frac{M_{\rm tot,\, xy}}{\langle M_{\rm fb,star}\rangle}.
\end{equation}
Both $A_{\rm tot,\,xy}$ and $A_{\rm max,\,xy}$ are dimensionless quantities expressed as hot phase mass per feedback-emitting stellar mass.

\subsubsection{Maximum evaporation efficiency}
Having measured $\mmh$, we can now determine the maximum evaporation efficiency factors, $A_{\rm max,\, xy}$, for the cold and warm phases. The Lagrangian nature of the tracer particles allows us to reconstruct the thermodynamic history of the gas contributing to $\mmh$. By tracing these particles back to their pre-SN thermal states, we can determine whether the hot gas originates primarily from the cold or warm phase. However, this measurement will be strongly dependent on the initial distribution of the gas. To account for this dependence, we model the contribution of gas originating from phase $x$ to the maximum hot-gas mass as
\begin{equation}\label{eq:m_max}
    M_{\rm{max,\,xy}} = f_{\rm x, max} \, \mmh
\end{equation}
where $f_{\rm x, max}$ is the fraction of $\mmh$ composed of gas that comes from phase $x$. This quantity depends on the amount of available material in the ambient medium. In particular, the amount of gas available for evaporation from phase $x$ is set by its initial mass fraction, $f_{\rm x,init}$, in the environment surrounding the SN explosion. 

\begin{figure}
    \centering
    \includegraphics[width=0.45\textwidth]{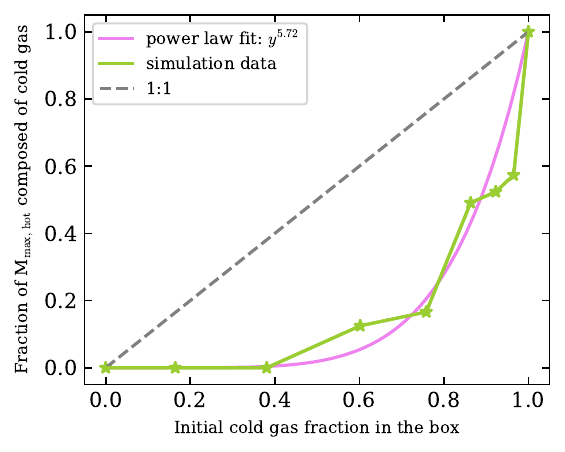}
    \caption{Relationship between the initial mass fraction of the cold phase, $f_{\rm cold,\, init}$, and the fraction of the peak hot-gas mass originating from that phase, $f_{\rm cold,\,max}$. The measurements (in green) are obtained from a turbulent simulation with $\nh=1.49\,{\rm cm^{-3}}$, in which the initial phase distributions are varied by changing the strength of the UV radiation field. The solid pink line shows the best-fitting relation, $f_{\rm cold,\, max}=f_{\rm cold,\,init}^{5.72}$, which we use to estimate the cold phase contributions to $\mmh$.} 
    \label{fig:fc_frac} 
\end{figure}

To quantify this dependence, we perform an additional series of turbulent-background simulations with fixed density, our fiducial $\nh = 1.49\, \rm cm^{-3}$, but different strengths of the UV radiation field, which modifies the initial phase distribution while leaving the overall setup unchanged. We show the resulting relationship between $f_{\rm x,\, init}$ and $f_{\rm x,\, max}$ for the cold phase in Figure \ref{fig:fc_frac}. By construction, $f_{\rm cold,\, init} = 0$ corresponds to an ambient medium in which none of $\mmh$ is made of cold gas, whereas $f_{\rm cold,\, init}=1$ corresponds to $\mmh$ being composed entirely of cold gas. We find the correlation between the fractions to be well described by $f_{\rm cold,\, max} = f_{\rm cold,\, init}^{5.72}$. The steepness of this relation indicates that the gas contributing to the peak hot-gas reservoir does not simply reflect the global, available phase fractions of the ambient medium. Instead, the SN preferentially expands through lower-density regions and therefore efficiently heats gas that is already warm, while largely avoiding dense cold structures. As a result, cold gas contributes significantly to $\mmh$ only when it constitutes a substantial fraction of the local environment and cannot be bypassed by the expanding remnant. We note, however, that this calibration is derived from the $\nh=1.49\, {\rm cm^{-3}}$ turbulent simulations, for which a suite of varying UV-field strengths was available; we assume that the same scaling provides a reasonable approximation across the remainder of our parameter space. We therefore use this relation to estimate the phase-resolved contributions to the peak hot-gas mass as $M_{\rm max,\, ch} = \mmh \times f_{\rm cold,\, init}^{5.72}$ for cold-to-hot transition and $M_{\rm max,\, wh} = \mmh \times (1 - f_{\rm cold,\, init}^{5.72})$ for warm-to-hot.

Finally, we can derive the final formula for the $A_{\rm max,\,xy}$ efficiency factor as a function of $f_{\rm x, init}$, $\nh$ and $\tcool$. By inserting Equation \ref{eq:mhot_max_turb} into Equation \ref{eq:m_max}, and combining it with the definition of the evaporation efficiency (Equation \ref{eq:A_max_xy}), we obtain
\begin{equation} \label{eq:A_max_combined}
A_{\rm max,\,xy} = 506\, f_{\rm x,\,max} \left(\frac{\nh}{1\,{\rm cm^{-3}}}\right)^{-0.30} \left(\frac{\tcool}{1\,{\rm Myr\, cm^{-3}}}\right)^{0.38},
\end{equation}
where the numerical prefactor follows from $1.05\times10^4\,M_\odot/\langle M_{\rm fb\,star}\rangle$, with $\langle M_{\rm fb\,star}\rangle=20.74\,M_\odot$ (see Equation \ref{eq:m_fb}). 

\begin{figure*}
    \centering 
    \includegraphics[trim=25mm 115mm 55mm 90mm, clip,width=0.95\textwidth]{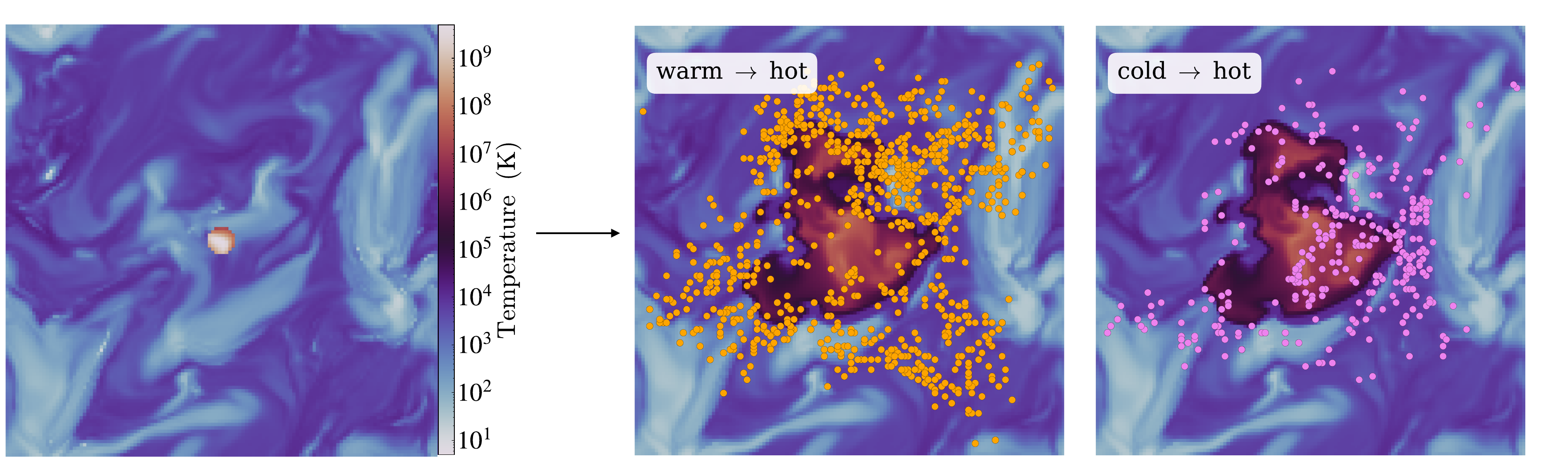}
    \caption{Spatial distribution of gas evaporation for a simulation with $\nh = 1.49\,\rm{cm^{-3}}$, $Z = 1 Z_{\odot}$ and a lowered UV field ($G_0 = 0.2$). The left panel shows the initial distribution of gas at $t = 10^2$ yr, with the color bar showing the gas temperature. The middle panel shows locations where tracer particles transition from the warm to the hot phase (orange points), while the right panel shows the cold-to-hot transitions (pink points), integrated over the full simulation duration (1 Myr) and projected along the $z$-axis. The background shows a gas temperature slice through the center of the box as the SN is reaching its maximum $\mmh$ at $t = 10^5$ yr. The points located outside the hot bubble show that evaporation into the hot phase continues even after the hot gas mass reaches its maximum. Warm gas, being more diffusely distributed throughout the simulation domain, leads to a more spatially uniform warm-to-hot evaporation pattern, whereas cold gas, concentrated in dense cloud structures, evaporates in more localized regions.} 
    \label{fig:evaporation_map} 
\end{figure*}

\subsubsection{Integrated evaporation efficiency.} 

\begin{figure*}
    \centering 
    \includegraphics[trim=0mm 5mm 0mm 0mm, clip,width=0.95\textwidth]{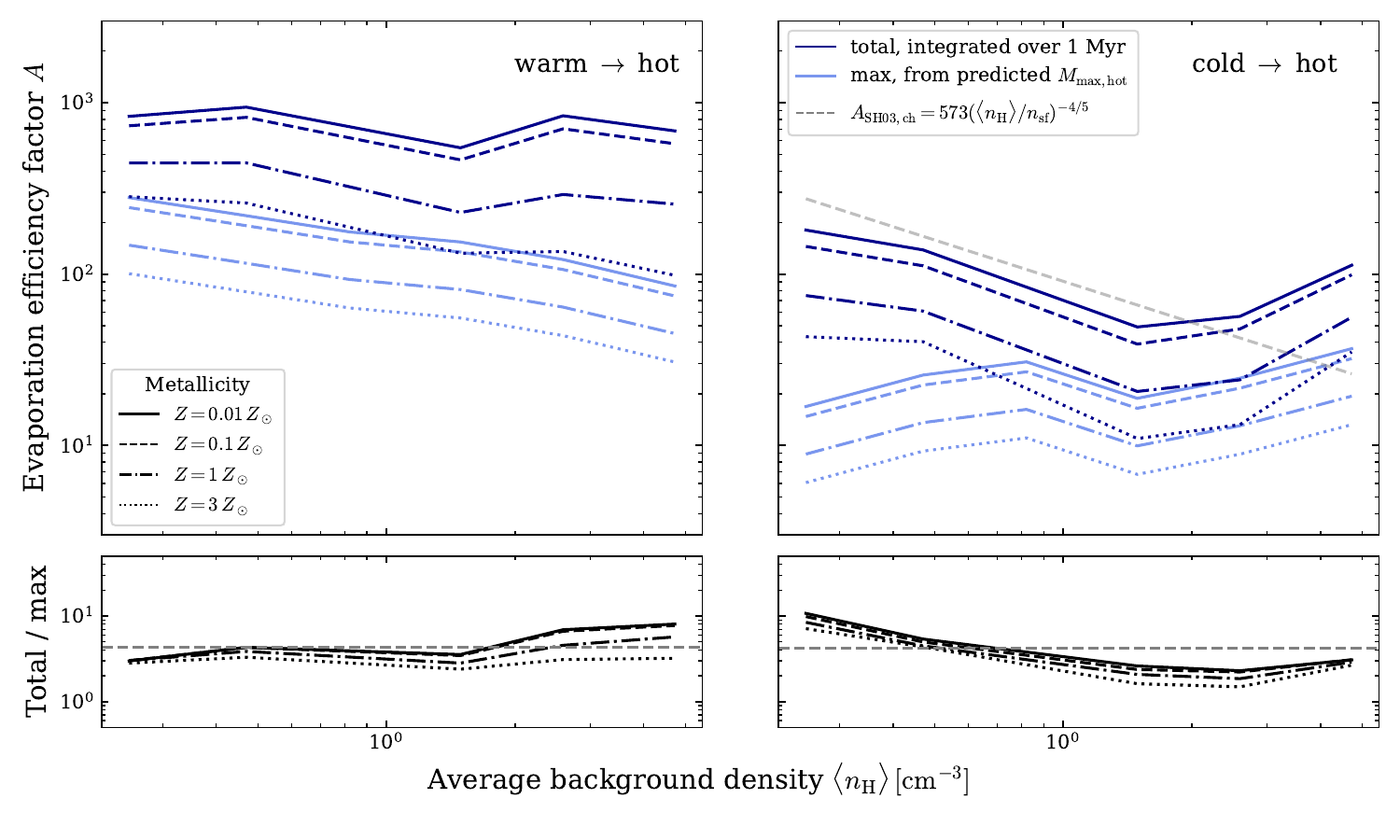}
    \caption{Total and maximum evaporation efficiencies as functions of density and metallicity in the turbulent simulations. The upper panels show the total evaporation efficiencies, $A_{\rm tot,\, xy}$, integrated over 1 Myr (dark blue lines), and the corresponding maximum efficiencies, $A_{\rm max,\, xy}$ (light blue lines) for the warm-to-hot (left column) and cold-to-hot (right column) transitions. The $A_{\rm max,\, xy}$ values are obtained from the predicted $\mmh$ (Equation \ref{eq:A_max_combined}). Different line styles denote the four metallicities indicated in the legend. The grey dashed line in the right panel shows the \citet{mckee_77} scaling adopted in \sh with normalization used in IllustrisTNG: $A_{\rm ch} = 573\, (\nh/n_{\rm sf})^{-4/5}$, where $n_{\rm sf} = 0.1\, \rm cm^{-3}$. The lower panels show the ratio $A_{\rm tot,\, xy}/A_{\rm max,\, xy}$, illustrating the contribution of evaporation occurring after the hot-gas mass reaches its maximum. The ratios show a modest dependence on metallicity and ambient density, but remain sufficiently consistent to motivate approximate conversion factors between the maximum and total evaporation efficiencies (grey dashed lines in bottom panels): $A_{\rm tot,\, ch}\approx 4.17 \,A_{\rm max,\, ch}$ and $A_{\rm tot,\, wh}\approx 4.28 \,A_{\rm max,\, wh}$.} 
    \label{fig:A_factors_both} 
\end{figure*}

While the maximum evaporation efficiency directly measures the amount of hot gas present at a given time, it does not capture the total amount of gas transferred into the hot phase over the duration of the simulation. Since the hot-phase mass evolves significantly following the SN explosion (see Figure \ref{fig:fractions_evolution}), we now focus on the total evaporation efficiency, which characterizes the entire hot-gas reservoir produced by a single SNa. This integrated quantity is particularly relevant for applications that model evaporation and cooling as separate processes, such as subgrid models of the multiphase ISM, where the cumulative mass processed through the hot phase can be more informative than the instantaneous hot-gas mass. To compute this quantity, we track the phase transitions with Lagrangian tracer particles. A tracer is counted as contributing to the evaporation from phase $x$ to phase $y$, and therefore $M_{\rm tot, \, xy}$, if it satisfies two conditions: \\
\noindent
\emph{(a)} it initially has a temperature corresponding to phase $x$ in the first snapshot of the simulation (pre-explosion), and \\
\noindent
\emph{(b)} it reaches a temperature corresponding to phase $y$ at least once between $t = 0$ and $t = 10^{6}$ yr after the SN explosion. 

In other words, each tracer is assigned to a single transition; for example, a particle that evolves directly from the cold phase to the hot phase is counted only as a cold-to-hot transition and not as a warm-to-hot transition.

To illustrate where gas evaporation occurs, we show in Figure \ref{fig:evaporation_map} the spatial distribution of phase transitions integrated over the duration of a single, turbulent-background simulation, with the initial conditions shown in the left panel. The middle panel displays warm-to-hot transitions (orange points), while the right panel shows cold-to-hot transitions (pink points). The SN remnant in the background is showed at $t = 10^5$ yr; therefore the points located outside of the bubble show evaporation occurring after the peak mass of the hot phase is reached by the expanding bubble. Evaporation from the warm phase occurs throughout a larger fraction of the simulation volume, reflecting the diffuse nature of the warm gas. In contrast, cold-to-hot transitions are concentrated closer to the dense cloud structures, producing a much more localized evaporation pattern. 

Having established the methodology for determining both $M_{\rm max,\,xy}$ and $M_{\rm tot,\,xy}$, and hence the corresponding evaporation efficiencies $A_{\rm max,\,xy}$ and $A_{\rm tot,\,xy}$, we now compare these quantities to identify how the total evaporated mass relates to the peak hot-gas reservoir. We evaluate Equation \ref{eq:A_max_combined} using the measured $f_{\rm x,\,init}$, $\nh$ and $\tcool$ from the turbulent simulations. This gives the predicted maximum evaporation efficiencies $A_{\rm max,\,wh}$ (left column) and $A_{\rm max,\,ch}$ (right column) shown in light blue in the upper panels of Figure \ref{fig:A_factors_both}. The dark blue curves show the corresponding measured total efficiencies, $A_{\rm tot,\,xy}$, obtained by measuring $M_{\rm tot,\, xy}$ with tracer particles in the same simulations. Different line styles correspond to the metallicities indicated in the legend.

The two evaporation channels exhibit distinct responses to the ambient density. The warm-to-hot efficiency, both $A_{\rm max,\,wh}$ and $A_{\rm tot,\,wh}$, decreases systematically with increasing density, whereas the cold-to-hot efficiency shows a weaker and more complex density dependence, depending on whether we consider the total or the maximum evaporation. This difference indicates that the two evaporation channels should not, in general, be expected to obey the same density scaling. The cold-to-hot efficiencies can also be compared with the evaporation prescription of \citet{mckee_77}, which is adopted in \sh and whose normalization is set to the IllustrisTNG value, $A_{\rm SH03,\, ch}=573(\nh/n_{\rm sf})^{-4/5}$, where $n_{\rm sf}=0.1\,{\rm cm}^{-3}$. This scaling is shown by the grey dashed line in the top right panel of Figure \ref{fig:A_factors_both}. While this scaling captures the qualitative expectation that the total cold-to-hot evaporation should become less efficient in denser environments, both our measured and predicted cold-to-hot efficiencies exhibit a substantially weaker density dependence than the relatively steep decline of the $-4/5$ power law, with the predicted $A_{\rm max,\,ch}$ \textit{increasing} weakly with density. This difference is important because the \citet{mckee_77} relation is derived for idealized evaporation (see the discussion in Section \ref{sec:efficiency}), whereas our simulations include turbulent mixing, radiative cooling, and the time-dependent evolution of the hot bubble.

Furthermore, the comparison between the light and dark blue lines show that $A_{\rm tot,\,xy}$ is systematically larger than $A_{\rm max,\,ch}$, as expected: the latter measures the amount of material incorporated into the hot phase up to the time at which $M_{\rm hot}$ reaches its maximum, whereas the former includes evaporation over the full duration of the simulation. The difference between the two therefore quantifies the additional evaporation that occurs after the hot-gas reservoir has reached its peak. The lower panels show the ratio between the measured $A_{\rm tot,\,xy}$ and the predicted $A_{\rm max,\,xy}$, which provides a direct measure of the amount of this post-maximum evaporation. Values close to unity indicate that most of the evaporated mass is incorporated into the hot phase before $M_{\rm hot}$ peaks, while larger values indicate a substantial contribution from evaporation occurring at later times.

The ratio exhibits a systematic dependence on metallicity. In both evaporation channels, the ratio tends to decrease as the metallicity increases, indicating that the total and maximum evaporation efficiencies become progressively more similar in more strongly cooling environments. In other words, enhanced radiative cooling reduces the amount of evaporation that occurs after the hot-gas reservoir reaches its maximum. This behavior is physically intuitive: stronger cooling limits the expansion and persistence of the hot bubble, reducing the time and volume available for additional evaporation after the peak in hot gas mass. Consequently, $A_{\rm tot,\,xy}$ approaches $A_{\rm max,\,xy}$ as the cooling becomes more efficient.

Moreover, for the warm-to-hot channel, the ratio shows a tendency to increase toward higher densities, indicating that an increasingly important fraction of the total warm-gas evaporation occurs after the hot-gas mass has peaked. On the other hand, the cold-to-hot ratio decreases with increasing the ambient density. We speculate that at higher ambient densities, cold gas is less efficiently evaporated after the hot-gas mass reaches its peak. Nevertheless, the overall variation in the ratios is modest enough that a simple conversion between the two quantities provides a useful approximation. We therefore take the density and metallicity average of the ratios (grey dashed lines in bottom panels) and write: $A_{\rm tot,\,ch} \approx 4.17 \,A_{\rm max,\,ch}$ and $A_{\rm tot,\,wh} \approx 4.28\,A_{\rm max,\,wh}$. By combining these relations with Equation \ref{eq:A_max_combined}, we can express the total SN-driven evaporation efficiency $A_{\rm tot,\, xy}$ directly in terms of $\tcool$, $\nh$, and $f_{\rm init}$. This allows the total SN-driven mass evaporated into the hot phase (up to 1 Myr) to be expressed in terms of the initial gas conditions, provided $\tcool$ is known, without explicitly following the full time evolution of gas evaporation. 

\section{Discussion} \label{sec:dis}

\begin{figure*}
    \centering
    \includegraphics[width=0.95\textwidth]{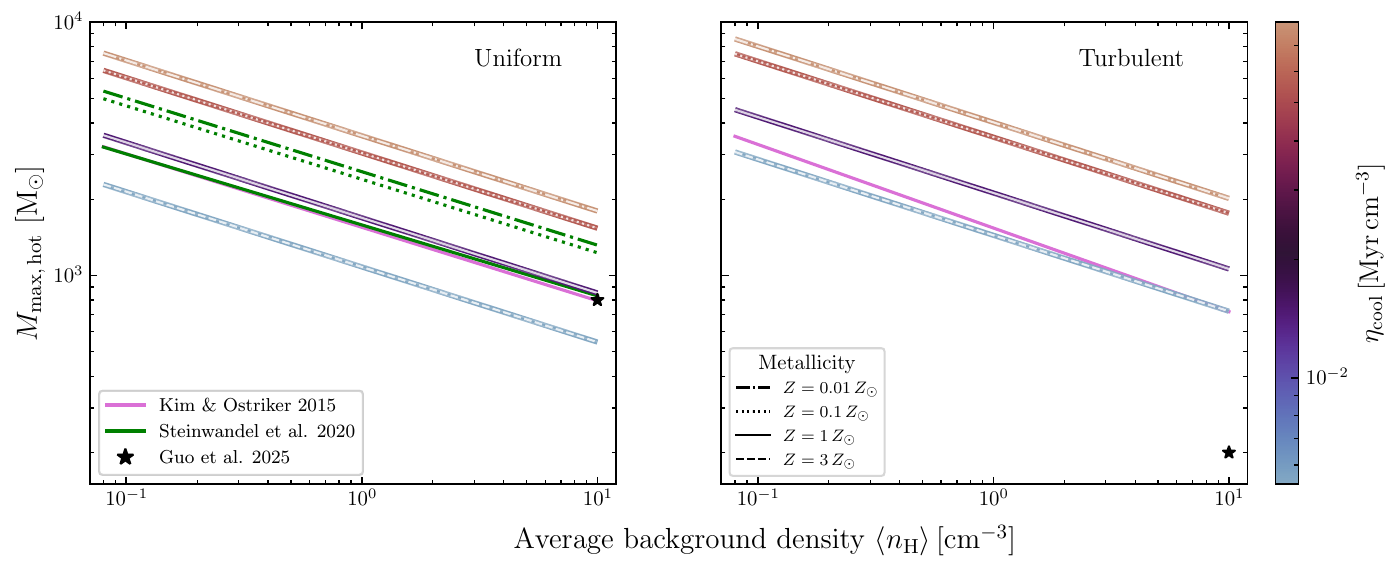}
    \caption{Maximum hot gas mass, $\mmh$, as a function of the mean background density for the derived scalings (Equations \ref{eq:mhot_max_uni} and \ref{eq:mhot_max_turb}). The lines are coloured by $\tcool$, the density-normalized integrated cooling time between $10^{4.5}\,\mathrm{K}$ and $10^{5.1}\,\mathrm{K}$. The left panel shows the uniform medium runs, while the right panel shows the turbulent background runs. Different line styles correspond to metallicities of $Z=3\,Z_\odot$ (dashed), $Z=Z_\odot$ (solid), $Z=0.1\,Z_\odot$ (dotted), and $Z=0.01\,Z_\odot$ (dash-dotted). The pink lines shows the scaling relation derived by \citet{kim_15}, $\mmh\approx 1550\,\nh^{-0.29}\,M_\odot$ for the uniform runs and $\mmh\approx 1540\,\nh^{-0.33}\,M_\odot$ for the turbulent ones. The green lines in the uniform panel show the relation from \citet{steinwandel_20}, for metallicities $Z=0.01$, $0.1$, and $1\,Z_\odot$ respectively: $\mmh\approx 2570\,\nh^{-0.28}\,M_\odot$, $\mmh\approx 2398\,\nh^{-0.29}\,M_\odot$ and $\mmh\approx 1584\,\nh^{-0.29}\,M_\odot$. The black star indicates the result of \citet{guo2024} shown for a total number density of $n=10\,\mathrm{cm^{-3}}$ in their work; for consistency with our simulations, we plot this result at $\nh=10\,\mathrm{cm^{-3}}$. Across the density range considered, the derived scaling reproduces the weak density dependence predicted by previous studies, with $M_{\rm max,hot}$ decreasing approximately as $\sim \nh^{-0.3}$. Variations in metallicity primarily shift the normalization through changes in the cooling time, while preserving a similar density scaling.} 
    \label{fig:summary_plot} 
\end{figure*}

\subsection{Comparison of $\mmh$ with previous work} \label{sec:dis_lit}

In Figure \ref{fig:summary_plot}, we compare our fitting formula for the peak hot-gas mass, $\mmh(\nh, \tcool)$, in coloured lines, with scaling relations from the literature. The colour indicates the cooling time, while the points show direct measurements from simulations with the fiducial cooling model (i.e., solar metallicity, $Z = 1 Z_\odot$). The left panel shows the uniform-medium scaling (Equation \ref{eq:mhot_max_uni}) alongside the relation of \citet{kim_15} in pink, whereas the right panel compares our turbulent-medium scaling (Equation \ref{eq:mhot_max_turb}) with their turbulent result. We additionally show the uniform-background relations derived by \citet{steinwandel_20} in green. 

For the solar-metallicity, uniform-medium case, our relation exhibits a density dependence that closely matches those found in previous studies. The turbulent simulations show the same qualitative trend, although with a somewhat shallower dependence on ambient density that reported by \citet{kim_15}. The various line styles denote different metallicities. Lower-metallicity runs systematically produce larger values of $\mmh$, owing to their longer cooling times. This behaviour is consistent with the interpretation that the peak hot-gas mass is primarily regulated by cooling in the $10^{4.5}$--$10^{5.1}\,\mathrm{K}$ temperature range. Differences between the various studies are therefore expected, as they employ different cooling prescriptions and consequently predict different cooling times in this regime.

The black star-shaped points in both panels correspond to the results of \citet{guo2024}: $200\,M_\odot$ for their turbulent/cloudy simulations and a value approximately four times higher for their uniform simulations. The authors attribute this factor-of-four difference to significant energy losses through radiative cooling at the shock--cloud interfaces in the turbulent case, which reduces the energy available to drive evaporation. Their results are given for a total number density of $n=10\,\mathrm{cm^{-3}}$; we plot these values at $\nh=10\,\mathrm{cm^{-3}}$. In contrast, we do not find a dependence of the maximum hot-gas mass on the structure of the ambient medium. Our fits give nearly identical normalizations for the uniform and turbulent simulations with similarly weak density dependences $\nh^{-0.29}$ and $\nh^{-0.30}$ (see Equations \ref{eq:mhot_max_uni} and \ref{eq:mhot_max_turb}). The dependence on the cooling time is also comparable, with exponents of $0.44$ and $0.38$ for the uniform and turbulent runs, respectively. This difference may reflect the different treatment of the turbulent background considered in the two studies; in particular, the magnitude of radiative losses at shock--cloud interfaces depends on the structure and properties of the multiphase medium and on how the interaction between the SN remnant and dense gas is resolved.

Additional discrepancies may arise from differences in the numerical setup, including the choice of hydrodynamic solver; the latter has been shown to affect the evolution of supernova remnants \citep{steinwandel_20}. Moreover, our numerical setup neglects thermal conduction, which can enhance the evaporation of cold material into the hot phase and modify the resulting hot-gas mass \citep{1977_cowie, 1977_mckee_cowie, 1981_cowie, steinwandel_20}.

\subsection{Analytical prediction of $\mmh$} \label{sec:analytical}

To assess whether the scaling relations derived from our numerical results can be physically understood from simple analytical considerations, here we derive an idealized estimate for the maximum mass of the hot phase as a function of gas density and radiative cooling efficiency. The main assumption is that the hot-gas mass grows as the SN remnant sweeps up ambient material, but eventually reaches a maximum once radiative cooling becomes important \citep{draine_11}. The swept-up hot mass can be approximated as 
\begin{equation}\label{eq:mhot_sedov}
M_{\rm hot} = \frac{4}{3}\pi R^3 \,\langle \rho \rangle,
\end{equation}
where $\langle \rho \rangle$ is the average density of the ambient medium and $R$ is the shock radius. The Sedov-Taylor solution \citep{1946_sedov}, which describes the adiabatic expansion of a blast wave into a uniform medium, shows this quantity evolves with time $t$ as
\begin{equation}
R = \left(\frac{E_0 t^2}{\langle \rho \rangle}\right)^{1/5},
\end{equation}
where $E_0$ is the total energy of the exploding SNe ($\approx 10^{51}$ erg). Substituting this relation into Equation \ref{eq:mhot_sedov} yields
\begin{equation}\label{eq:mhot_t}
M_{\rm hot} = \frac{4}{3}\pi \langle \rho \rangle^{2/5} E_0^{3/5} t^{6/5}.
\end{equation}
To determine when the growth of the hot phase stalls, we have to estimate the cooling time of the SN-heated gas.\footnote{For simplicity, we assume a uniform temperature and density when estimating the cooling time. A more accurate analytic treatment would account for the radial temperature and density profiles of the Sedov--Taylor self-similar solution when evaluating the cooling time, and would also refine the enclosed swept-up mass in Equation \ref{eq:mhot_sedov} accordingly. An example of such a derivation is given in \citet[\S\,39.1.2]{draine_11}.} To that end, we assume that the explosion energy $E_0$ is distributed throughout the hot bubble and tie $E_0$ to the average temperature inside the bubble $\langle T_{\rm SN} \rangle$ as
\begin{equation}
E_0 \approx \frac{3}{2} N_{\rm hot} k_{\rm b} \langle T_{\rm SN} \rangle,
\end{equation}
where $N_{\rm hot}=M_{\rm hot}/\bar{m}$ and $\bar{m}$ is the mean particle mass. This relation is approximate, as in practice the explosion energy is distributed between the thermal energy of the hot gas and the kinetic energy of the expanding shock. Here, we use it as an order-of-magnitude estimate relating the characteristic thermal energy of the hot bubble to the initial explosion energy. Rearranging gives
\begin{equation} \label{eq:tav}
\langle T_{\rm SN} \rangle = \frac{2}{3} \frac{E_0\bar{m}} {k_{\rm b}M_{\rm hot}}.
\end{equation}
For simplicity, we adopt the isochoric cooling time defined as
\begin{equation} \label{eq:tcool}
\tau_{\rm cool} = \frac{3}{2} \frac{k_{\rm b}\langle T_{\rm SN} \rangle} {\langle n \rangle\mathcal{L}} \,.
\end{equation}
Here, $\tau_{\rm cool}$ represents the same physical cooling timescale as $t_{\rm cool}$ introduced above, namely the timescale over which the gas loses a significant fraction of its thermal energy. There are, however, two approximations that we make here; first, rather than integrating the cooling time over the temperature interval $T_{\rm min}$--$T_{\rm max}$ as in our numerical calculation (see Equation \ref{eq:tcool_def}), we evaluate the cooling time at a specific temperature --- the instantaneous temperature of the remnant, which changes with time (or equivalently with radius) as the supernova remnant expands. Second, we approximate the cooling function as a power law,
\begin{equation}
\mathcal{L}(T) = \mathcal{L}_0 \left(\frac{\langle T_{\rm SN} \rangle}{T_0}\right)^{-\beta}.
\end{equation}
While a power-law approximation is generally not an accurate representation of the cooling function over a broad temperature range, it provides a reasonable approximation over the temperature range characteristic of the hot gas considered in this simplified derivation. We substitute this relation and Equation \ref{eq:tav} into Equation \ref{eq:tcool}, and using $\langle n \rangle = \langle \rho \rangle/\bar{m}$, we obtain 
\begin{equation}\label{eq:tcool_mhot}
\begin{split}
\tau_{\rm cool}
&=
\frac{3}{2}
\frac{k_{\rm b} \bar{m}}
{\langle \rho \rangle \mathcal{L}_0 T_0^{\beta}}
\left(
\frac{2}{3}
\frac{E_0 \bar{m}}
{k_{\rm b}}
\right)^{1+\beta}
M_{\rm hot}^{-(1+\beta)}
\\
&=
\left(\frac{2}{3 T_0 k_{\rm b}}\right)^{\beta}
\frac{E_0^{1+\beta} \bar{m}^{2+\beta}}
{\langle \rho \rangle\mathcal{L}_0}
M_{\rm hot}^{-(1+\beta)}.
\end{split}
\end{equation}
We assume that the hot phase reaches its maximum mass when radiative losses begin to significantly affect the remnant evolution at the end of the ST stage. We approximate the end of the ST stage as occurring at $t \approx \frac{1}{2}\tau_{\rm cool}$. The factor of $1/2$ is not physically unique, but is intended to indicate that cooling has begun while the gas has not yet undergone significant radiative losses; choosing a factor of order unity, such as $1$ or $1/3$, would be equally reasonable. By substituting Equation \ref{eq:tcool_mhot} into Equation \ref{eq:mhot_t}, we obtain
\begin{equation}
\begin{split}
M_{\rm hot} 
&=
\frac{4\pi}{3}\,2^{-6/5}
\left(\frac{2}{3T_0k_{\rm b}}\right)^{6\beta/5}
\\
&\quad \times
\mathcal{L}_0^{-6/5}
\bar m^{6(2+\beta)/5}
E_0^{(9+6\beta)/5}
\langle \rho \rangle^{-4/5}
M_{\rm hot}^{-6(1+\beta)/5}.
\end{split}
\end{equation}
By rearranging, we obtain the final expression
\begin{equation}
\begin{split}
M_{\rm hot}
&=
\Bigg[
\frac{4\pi}{3}\,
2^{-6/5}
\left(\frac{2}{3T_0 k_{\rm b}}\right)^{6\beta/5}
\\
&\qquad\qquad\times
\mathcal{L}_0^{-6/5}
\bar m^{6(2+\beta)/5}
E_0^{(9+6\beta)/5}
\langle \rho \rangle^{-4/5}
\Bigg]^{\frac{5}{11+6\beta}} \,.
\end{split}
\end{equation}
For a representative cooling-law slope of $\beta=0.7$, this expression reduces to $M_{\rm hot} \propto n ^{-0.26} \mathcal{L}_0^{-0.39}$. Since the cooling normalization scales with the density-independent integrated cooling time $\tcool$ as $\mathcal{L}_0 \propto \tcool^{-1}$ (see Section \ref{sec:max_mhot_cool}), the final prediction becomes $M_{\rm hot} \propto \langle n \rangle ^{-0.26} \tcool^{0.39}$. The analytical model therefore predicts that the maximum hot-gas mass should increase with cooling time and decrease with ambient density. Both the obtained cooling-time and density dependence are remarkably close to the exponents we measured in our simulations: $\nh^{-\alpha}$, with $\alpha$ in the range from $-0.29$ to $-0.30$ and $\tcool^{\beta}$, with $\beta$ in the range from $0.38$ to $0.44$ (see Equations \ref{eq:mhot_max_uni} and \ref{eq:mhot_max_turb}). 

\begin{figure*}
    \centering 
    \includegraphics[trim=20mm 160mm 205mm 60mm, clip, width=0.85\textwidth]{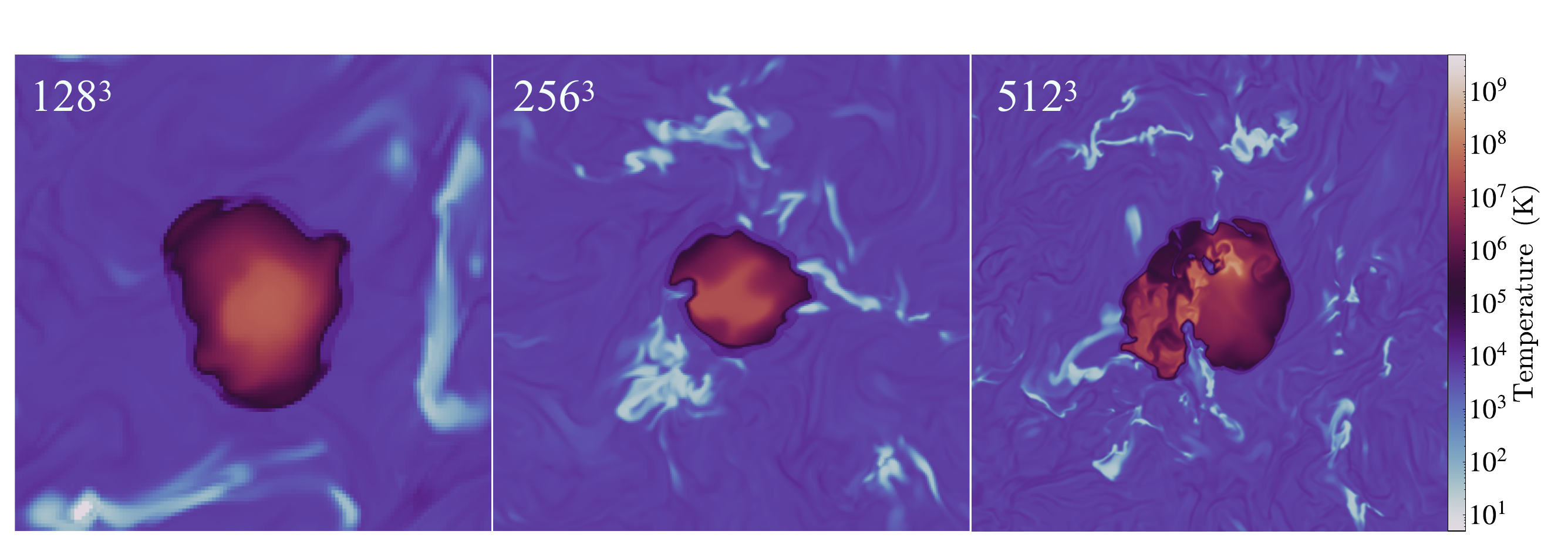}
    \caption{Gas temperature slices through the center of the box inside our turbulent-background SNe simulations at $t = 10^5$ yr, with average hydrogen number density $\nh = 1.49 \, \rm cm^{-3}$, gas metallicity $Z\,= 1\,Z_{\odot}$ and number of cells as labelled. Increasing the resolution from $128^3$ to $512^3$ cells results in a maximum $20\%$ difference in the measured $\mmh$. The difference in the morphology of the SNR is due to the increasingly fine filamentary and clumpy structures at higher resolution, which reflect the improved resolution of the turbulent background.}
    \label{fig:resolution_slices}
\end{figure*}

\subsection{Sensitivity to resolution, hydrodynamical
methods, explosion location and background pressure} \label{sec:reso}

To check the robustness of our results, we conducted a few tests with higher resolution for both the uniform and turbulent simulations. In addition to the simulations described in Section \ref{sec:meth}, performed at a resolution of $128^3$, we carried out selected simulations at $256^3$ and $512^3$. For the uniform runs, the resulting maximum hot gas mass differs by less than $\sim10\%$ across the explored density range, $\nh=0.1-5\,\mathrm{cm^{-3}}$. For the turbulent runs, at a fixed density of $1.49\,\mathrm{cm^{-3}}$, the maximum hot gas mass varies by approximately 15--20\% relative to the fiducial-resolution run, with no systematic trend with increasing resolution. The corresponding temperature slices for the turbulent runs are shown in Figure \ref{fig:resolution_slices}, illustrating the increasingly resolved structure of the SN remnant with increasing spatial resolution. While the detailed morphology of the bubbles differs strongly between resolutions, these differences primarily reflect the resolution-dependent structure of the turbulent background. Higher resolution resolves progressively smaller-scale structures, resulting in a greater abundance of small, dense clumps and finer filamentary features. Nevertheless, the overall filamentary and clumpy morphology is qualitatively preserved across resolutions and consistent with previous studies of turbulent interstellar clouds \citep{2010_audit, bellomi_20, godard_23, MalamudEtAl-2026}.

We also tested the sensitivity of our results to the choice of Riemann solver by repeating the uniform and turbulent simulations with background density $\nh = 1.49\, \rm cm^{-3}$ using both the exact solver (with CG85 reconstruction) and the HLLC solver (with linear reconstruction). For the turbulent case, the maximum hot gas mass differs by only $\sim2\%$ between the two solvers, while for the uniform case the difference is $\sim5\%$. 

Additionally, for the turbulent runs with the fiducial cooling model, we tested the sensitivity of our results to the explosion location by placing the supernova at fractional box coordinates $(0.25,\,0.25,\,0.25)$, $(0.5,\,0.5,\,0.5)$ (fiducial), and $(0.75,\,0.75,\,0.75)$. Since our analysis employs Monte Carlo tracer particles, a small level of run-to-run variation is expected by construction due to the stochastic nature of the tracer sampling. Across the three ambient densities considered ($\nh = 0.25,\ 1.49,\ 4.73\, \mathrm{cm^{-3}}$), the final amount differs by at most $\sim7\%$, $\sim5\%$, and $\sim15\%$, respectively, relative to the fiducial case. The largest deviation occurs in the highest-density simulation, where the supernova happens to explode within a dense, cold cloud. Overall, these tests indicate that our results are robust to the choice of explosion location.

We have also tested the sensitivity of the SN-driven hot gas mass to the assumed initial background temperature by repeating the key simulations with a uniform background density and varying the average temperatures. We find that the resulting $\mmh$ values show only a weak dependence on the initial thermal state of the gas. For $\nh=0.25\,\mathrm{cm^{-3}}$, varying the background temperature from the fiducial value of $T=8\times10^3$ K to $T=10^5$ K and $T=4\times10^3$ K changes $\mmh$ by only $\sim5\%$ and $\sim8\%$, respectively; even for a temperature variation spanning more than an order of magnitude, the resulting hot gas mass changes by less than $\sim10\%$. The largest variation in $\mmh$ occurs at the highest density: for $\nh=4.73\, \mathrm{cm^{-3}}$, reducing the background temperature to $T=10^3$ K decreases $\mmh$ by approximately $13\%$. This stronger sensitivity at high density is expected, as the increased ambient pressure and shorter cooling times can affect the evolution of the SN remnant and alter the amount of gas that can be maintained in the hot phase. Nevertheless, these differences remain subdominant compared to the impact of the background density and cooling strength, indicating that the SN-driven hot phase is largely insensitive to the assumed initial temperature across the explored parameter range.

\section{Conclusions} \label{sec:con}

In this work, we have provided a physically motivated characterization of feedback-driven phase transitions in a realistic multiphase ISM by explicitly linking the origin of the hot gas to the properties of the ambient medium and radiative gas cooling. Our main findings can be summarized as follows:

\begin{itemize}

    \item We show that the maximum mass of the hot ($T > 2 \times 10^4$ K) phase, $M_{\rm max,hot}$, is determined by the combined effect of ambient density and radiative cooling (Figure \ref{fig:combined_sn_cooling}), as increasing either the gas density or the cooling strength suppresses the build-up of hot gas (Figure \ref{fig:fractions_evolution}). Specifically, we find that the strongest predictor of $M_{\rm max,hot}$ is $\tcool$, the density-independent integrated cooling time of gas in the temperature range $10^{4.5}\,{\rm K} < T < 10^{5.1}\,{\rm K}$ (Figure \ref{fig:triangles}). This temperature range coincides with the regime that dominates the radiative cooling losses of the hot phase (Figure \ref{fig:dEdtdT}). Our results therefore suggest that the peak hot-gas mass is primarily regulated by the loss of energy --- stronger cooling leads to a smaller, less energetic shock bubble that stalls earlier, causing a lower amount of mass being transferred to the hotter phases.
    
    \item  We find that the maximum hot-gas mass follows similar scaling relations in both uniform and turbulent environments, with an approximately $\nh^{-0.3}$ dependence on the mean background density and a $\tcool^{0.4}$ dependence on the cooling time (see Equations \ref{eq:mhot_max_uni} and \ref{eq:mhot_max_turb}, as well as Figure \ref{fig:Mhot_tcool}). The similarity of these relations indicates that the large-scale structure of the ambient medium, whether uniform or turbulent, has only a minor impact on the maximum hot-gas mass. The obtained density dependence is consistent with previous numerical studies, despite the substantially different cooling prescriptions and ISM structures considered in this work (see Figure \ref{fig:summary_plot}).

    \item We show that the gas contributing to the maximum hot phase mass is predominantly drawn from the warm ISM --- the contribution from cold gas depends strongly on its initial abundance and follows $f_{\rm cold,max}\propto f_{\rm cold,init}^{5.72}$ (Figure \ref{fig:fc_frac}). This behaviour indicates that supernova remnants preferentially expand through low-density channels and efficiently heat diffuse warm material while largely avoiding dense cold structures (Figure \ref{fig:evaporation_map}).

    \item We distinguish between the peak hot-gas reservoir and the cumulative amount of hot gas produced by the supernova remnant over 1 Myr. We then define the corresponding peak and integrated efficiencies of SN-driven phase transitions, $A_{\rm max,xy}$ and $A_{\rm tot,xy}$ respectively, that quantify the amount of gas transferred from phase $x$ to phase $y$ per unit mass of feedback-emitting stars. 

    \item We find distinct density dependencies for the two evaporation channels (see Figure \ref{fig:A_factors_both}): the warm-to-hot efficiency decreases with increasing ambient density, while the cold-to-hot efficiency exhibits a weaker density dependence, which varies between $A_{\rm max,ch}$ and $A_{\rm tot,ch}$. In particular, the cold-to-hot efficiencies do not follow the commonly adopted \citet{mckee_77} scaling, $A_{\rm ch}\propto\nh^{-0.8}$, over the density range explored here. These results suggest that the density dependence of evaporation in turbulent, radiatively cooling gas differs from that assumed in commonly adopted sub-grid feedback prescriptions.

    \item We find that the total and maximum evaporation efficiencies are related through approximately constant conversion factors, $A_{\rm tot,\, ch}\approx 4.17 \,A_{\rm max,\, ch}$ and $A_{\rm tot,\, wh}\approx 4.28 \,A_{\rm max,\, wh}$, with only modest variations across the range of densities and metallicities explored. This indicates that a substantial fraction of the evaporated mass is incorporated into the hot phase after the hot-gas reservoir reaches its maximum, and therefore estimates based solely on the peak hot-gas mass systematically underestimate the total mass transferred into the hot phase. 

\end{itemize}

We note that our results represent idealized setups in which we model an individual supernova explosion in a turbulent gas volume. However, in the ISM, such explosions rarely take place in isolation -- the youngest and largest stars inside molecular clouds tend to explode in clusters, which results in multiple SNe bubbles overlapping each other and collective feedback effects \citep{martizzi_15}. Addressing this limitation will require future studies to explore multiple, spatially correlated supernova events within the same turbulent volume \citep[see e.g.,][]{kim_15}, enabling a more complete understanding of how successive explosions shape gas evaporation, mixing, and phase transitions in the ISM. In addition, a more complete treatment of supernova remnant evolution would require exploring the role of reverse shocks, which may influence the long-term evolution of hot gas phases in the ISM \citep{truelove_evolution_1999}.

Nevertheless, the phase-transition efficiencies presented here provide key ingredients for subgrid SN feedback models in cosmological simulations, where the exchange of mass between cold, warm, and hot gas cannot be resolved directly. By distinguishing between the peak hot-gas reservoir and the cumulative hot phase mass, our framework captures both the instantaneous and integrated impact of SN-driven evaporation in the turbulent ISM.

\begin{acknowledgments}

This research was supported in part by the National Science Foundation under grant No. AST 2338388. ZK acknowledges the support from the Fulbright Graduate Award awarded by the Polish-U.S. Fulbright Commission (Scholarship Agreement No. PL/2023/2/GS). SB acknowledges support from the ISF grant number 2071540, the GIF grant number I-1568-303.7/2024, the NSF-BSF grant number 2023761, and the Alon Fellowship prize for junior faculty.
Support for VS was provided by Harvard University through the Institute for Theory and Computation Fellowship. The simulations presented in this paper were carried out on the ZARATAN cluster at University of Maryland. Analyses presented in this paper were greatly aided by the following free software packages: {\tt yt} \citep{yt}, {\tt NumPy} \citep{numpy_ndarray}, {\tt SciPy} \citep{scipy}, and {\tt Matplotlib} \citep{matplotlib}. We have also used the Astrophysics Data Service (\href{http://adsabs.harvard.edu/abstract_service.html}{ADS}) and \href{https://arxiv.org}{arXiv} preprint repository extensively during this project and writing of the paper.

\section{contribution}

Using the CRediT (Contribution Roles Taxonomy) system (\url{https://authorservices.wiley.com/author-resources/Journal-Authors/open-access/credit.html}), the main roles of the authors were:\\
\noindent
\textbf{ZK}: conceptualization, methodology, data curation, resources, formal analysis, investigation, visualization, writing – original draft.\\
\textbf{BD}: conceptualization, methodology, supervision, validation, writing – review \& editing. \\
\textbf{VS}: conceptualization, methodology, supervision, resources, validation, writing – review \& editing. \\
\textbf{SB}: conceptualization, methodology, supervision, resources, validation, writing – review \& editing.\\
\textbf{UM}: resources, writing – review \& editing.

\end{acknowledgments}

\appendix

In Table \ref{tab:cooling_models}, we specify which cooling processes from the \citet{ploeckinger_20} tables were included in the cooling models shown in the left panel of Figure \ref{fig:cooling_models}.

\begin{table}

\centering
\caption{Cooling processes included in each model showed in the left panel of Figure \ref{fig:cooling_models}. Model M1 is our fiducial cooling model and corresponds to the full \citet{ploeckinger_20} cooling table.}
\label{tab:cooling_models}

\scriptsize
\setlength{\tabcolsep}{3pt}

\begin{tabular}{lccccccccccc}
\toprule
Process &
M1 & M2 & M3 & M4 & M5 & M6 &
M7 & M8 & M9 & M10 & M11 \\
\midrule
H                         & x &   & x & x & x & x & x & x & x & x & x \\
He                        & x &   & x & x & x & x & x &   & x & x & x \\
C                         & x & x & x & x & x & x & x & x & x & x & x \\
N                         & x &   & x & x & x & x &   & x &   & x & x \\
O                         & x & x & x & x & x & x &   &   &   & x &   \\
Ne                        & x &   & x & x & x &   &   &   &   &   & x \\
Mg                        & x &   &   &   & x &   &   &   &   &   & x \\
Si                        & x &   &   &   & x &   &   &   &   &   & x \\
S                         & x &   &   &   & x &   &   &   &   & x & x \\
Ca                        & x &   &   &   & x &   &   &   &   & x & x \\
Fe                        & x &   &   &   &   &   &   &   &   & x & x \\
Other metals              & x &   &   &   &   & x &   &   &   & x & x \\
H$_2$                     & x &   &   & x &   & x &   &   &   & x & x \\
Other molecules           & x &   &   & x &   & x &   &   &   & x & x \\
HD rotational cooling     & x &   &   & x &   & x &   &   &   & x & x \\
Brems. (H, He)            & x &   &   & x &   & x &   &   &   & x & x \\
Brems. (metals)           & x &   &   & x & x & x &   &   &   & x & x \\
Brems. ($e$--$e$)         & x & x &   & x & x & x &   &   & x & x & x \\
Compton                   & x & x &   & x & x & x &   &   & x & x & x \\
Dust                      & x & x &   & x & x & x &   &   & x & x & x \\
\bottomrule
\end{tabular}
\end{table}

\clearpage

\bibliographystyle{aasjournal}
\bibliography{SN_paper_old, SN_paper_new}

\end{document}